\documentclass[10pt,a4paper]{article}
\usepackage[utf8]{inputenc}
\usepackage[english]{babel}
\usepackage{amsmath}
\usepackage{amsfonts}
\usepackage{amssymb}
\usepackage{graphicx}
\usepackage[left=3cm,right=3cm,top=2cm,bottom=2cm]{geometry}
\usepackage{color}
\usepackage[dvipsnames]{xcolor}
\usepackage{floatflt}
\usepackage{fixltx2e}
\usepackage{wrapfig}
\usepackage[font=small, labelfont={sf,bf}, margin=0.5cm]{caption}
\usepackage[backend=bibtex,sorting=none,style=numeric-comp]{biblatex}
\usepackage{authblk} 

\title{\bf Fast-slow bursters in the unfolding of a high codimension singularity and the ultra-slow transitions of classes.\vspace{1ex}}

\author{Maria Luisa Saggio\thanks{maria-luisa.saggio@etu.univ-amu.fr}
}
\author{Andreas Spiegler}
\author{Christophe Bernard}
\author{Viktor K. Jirsa\thanks{viktor.jirsa@univ-amu.fr}
}

\affil{\textit{Institut de Neurosciences des Syst\`emes - Inserm UMR1106, Aix-Marseille Universit\'e, Marseille, France}}

\date{\small{\today}\vspace{-3ex}}

\begin{document}

\maketitle

%

\section*{Abstract}

Bursting is a phenomenon found in a variety of physical and biological systems. For example, in neuroscience, bursting is believed to play a key role in the way information is transferred in the nervous system. In this work, we propose a model that, appropriately tuned, can display several types of bursting behaviors. The model contains two subsystems acting at different timescales. For the fast subsystem we use the planar unfolding of a high codimension singularity. In its bifurcation diagram, we locate paths that underly the right sequence of bifurcations necessary for bursting. The slow subsystem steers the fast one back and forth along these paths leading to bursting behavior. The model is able to produce almost all the classes of bursting predicted for systems with a planar fast subsystems. Transitions between classes can be obtained through an ultra-slow modulation of the model's parameters. A detailed exploration of the parameter space allows predicting possible transitions. This provides a single framework to understand the coexistence of diverse bursting patterns in physical and biological systems or in models. 

\section{Introduction}

\begin{figure}[t]
\begin{minipage}{0.55\textwidth}
\includegraphics[width=\textwidth]{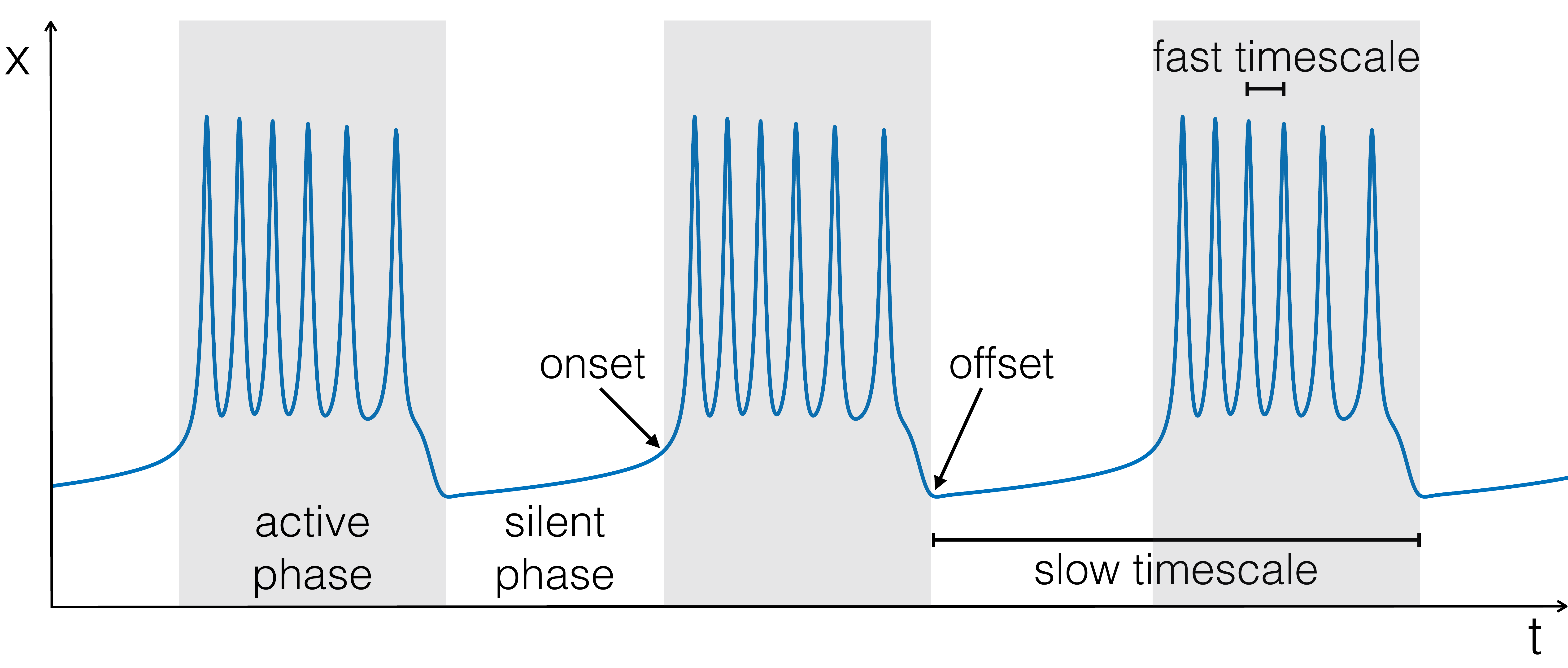} 
\end{minipage}
\begin{minipage}{0.45\textwidth}
\caption{\textbf{Bursting activity.} Bursters are characterized by the alternation between active (grey boxes) and silent (white) phases. In fast-slow bursters we can distinguish two rhythms, one within the active phase, the fast timescale, and the other between active and silent phases, the slow timescale. Oscillations start at the onset of the active phase and terminates at its offset.}
\label{fig_timescalesep}
\end{minipage}
\end{figure}

Many systems in nature can display bursts of activity that alternate with silent behavior \cite{rinzel1987formal,atwater1979nature}. An example of bursting is shown in Fig.~\ref{fig_timescalesep}. Bursting is in fact part of the dynamical repertoire of many chemical and biological systems and is the primary mode of electrical activity in several neurons and endocrine cells \cite{bertram1995topological,deschenes1982thalamic,crunelli1987ventral,wong1981afterpotential,harris1987multiple,johnson1992burst,dean1970glucose,ashcroft1989electrophysiology,hudson1979experimental}. Neuronal bursting, in particular, is of key importance for the production of motor, sensory and cognitive behavior \cite{fox2015bursting}. Bursts of activity are central to information processing, as they produce reliable synaptic transmission and as they can facilitate synaptic plasticity  \cite{lisman1997bursts}. Bursting can also be pathological. For example, epileptiform discharges are associated with bursts of neural ensembles with highly synchronized activity \cite{connors1984initiation}.

Modeling bursting behavior can help to uncover the mechanisms underlying the bursting dynamics in complex systems. Moreover, modeling gives the opportunity to perform in silico experiments to predict the outcome of manipulations of the system. For example, the Epileptor, which is a phenomenological model \cite{jirsa2014nature} for the most common bursting behavior in epilepsy, has been used to predict seizure propagation and recruitment in highly personalized virtual epileptic brains \cite{proix2016individual}. Different treatment strategies can be tested in silico in these virtual epileptic patients, such as interventions on the network topology, stimulations and parameters changes, providing a tool throughout the presurgical evaluation.

Bursting activities, though, can present large differences, such as differences in amplitude and frequency. Different properties at the onset and offset of the burst (i.e. active phase) have been linked to specific qualitative changes in the dynamics, which correspond to bifurcations occurring in a subsystem of the dynamical system \cite{rinzel1987formal,izhikevich2000neural}. Izhikevich used the onset/offset bifurcations pair criterion to compile a taxonomy of possible bursting classes \cite{izhikevich2000neural}. In the present study we provide a single autonomous model, comprising a minimal number of variables and parameters, able to produce many classes from this taxonomy. For this purpose, we make use of (i) the `dissection' method developed by Rinzel \cite{rinzel1987dissection} for the study of fast-slow bursters, namely bursters for which there is a timesecale separation between the rhythm of oscillation within the active phase and the rhythm at which silent and active phases alternate; (ii) the unfolding theory approach proposed by Golubitsky et al. \cite{golubitsky2001unfolding}, based on the idea that the bifurcations involved in bursting activity can be `collapsed to a single local bifurcation, generally of higher codimension'.

In Section \ref{sec_modeling}, we will briefly recall both the dissection method and the unfolding theory approach. We will also introduce a codimension 3 singularity, the degenerate Takens-Bogdanov (codim-3 deg. TB) singularity.  In Section \ref{sec_results} we will extend the unfolding approach to the deg. TB singularity and show how this allows for a rich repertoire of bursting classes. The model in fact is able to display almost all types of bursting behavior that have been predicted for systems with timescale separation and a planar subsystem acting on the fast timescale \cite{izhikevich2000neural}. We will explain in detail how to build the different classes of bursters and, furthermore, how to obtain transitions among classes with an ultra-slow modulation of the model parameters. In addition, we will show additional bursting classes obtained when varying a fourth parameter of the model. Finally, we will apply a measure for complexity based on codimensions \cite{golubitsky2001unfolding} to the bursting classes found in the model. This can help to understand the occurrence of bursting phenomena, in empirical data and models.

\section{Modeling fast-slow bursters}
\label{sec_modeling}

\subsection{Dissection method}
At least two rhythms characterize a burster: the rhythm of the oscillations within the active phase, and the rhythm of the alternation between active and silent phases (Fig.~\ref{fig_timescalesep}).

If the timescales of these two rhythms are sufficiently apart ($f_{fast} >10\, f_{\rm slow}$), we have a \textit{fast-slow} burster. Rinzel \cite{rinzel1985bursting} took advantage of this separation to analyze bursting in the Chay-Keyzer model for pancreatic $\beta$ cells. He applied a powerful method, called  `dissection', that is at the base of our work. The idea behind this method is that we can distinguish two subsystems, the slow and the fast ones, operating at $f_{\rm fast}$ and $f_{\rm slow}$ respectively, and that the variables of the slow subsystem enter the fast subsystem's equation as parameters.

The fast-slow burster can thus be described by
\begin{equation}
\begin{cases}
\mathbf{\dot{x}}=f(\mathbf{x},\mathbf{z})\\
\mathbf{\dot{z}}=c\,g(\mathbf{x},\mathbf{z})
\end{cases}
\end{equation}
where $\mathbf{x}\in\mathbb{R}^n$ is the state vector of fast variables, $\mathbf{z}\in\mathbb{R}^m$ is the vector of slow ones and $c=1/\tau$ is the inverse of the characteristic time constant of the separation between the two rhythms.

The fast subsystem can be analyzed isolated from the slow one. One can thus build a bifurcation diagram showing how the state space topology of the fast subsystem changes for different values of the slow variables $\mathbf{z}$, here playing the role of bifurcation parameters. If the timescale separation holds, the coupling with the slowly changing $\mathbf{z}$ moves the fast subsystem in this bifurcation diagram, without affecting the topology of the latter.

\subsection{Classification of bursters}

When coupled together, the two subsystems must fulfill at least two requirements to produce bursting activity. First, the fast subsystem should be able to display both silent and oscillatory activity depending on the value of its parameters, that is the slow variables \cite{de1998multiple}. This implies that the dimensionality of the fast subsystem should be $n\geqslant2$, to allow for the existence of a limit cycle. Second, the slow subsystem should oscillate to promote the switching between silence and fast oscillations in the fast subsystem. This, though, does not necessarily require a bidimensional slow subsystem. The slow oscillation, in fact, can occur through two mechanisms \cite{izhikevich2000neural}: 
\begin{itemize}
\item \textit{Slow-wave burster} - The slow subsystem is a self-sustained oscillator, thus feedback from the fast to the slow subsystem is not required. In this case, the slow subsystem must be at least two-dimensional $m\geqslant2$.
\item \textit{Hysteresis-loop burster} - The slow subsystem oscillates due to feedback from the fast subsystem. This can occur if the fast subsystem shows hysteresis between the silent and active states, which can be used to inform the slow subsystem about the state of the fast subsystem (e.g., by baseline). In this case one slow variable is enough, $m\geqslant1$.
\end{itemize}

\begin{figure}[]
\center
\includegraphics[width=0.93\textwidth]{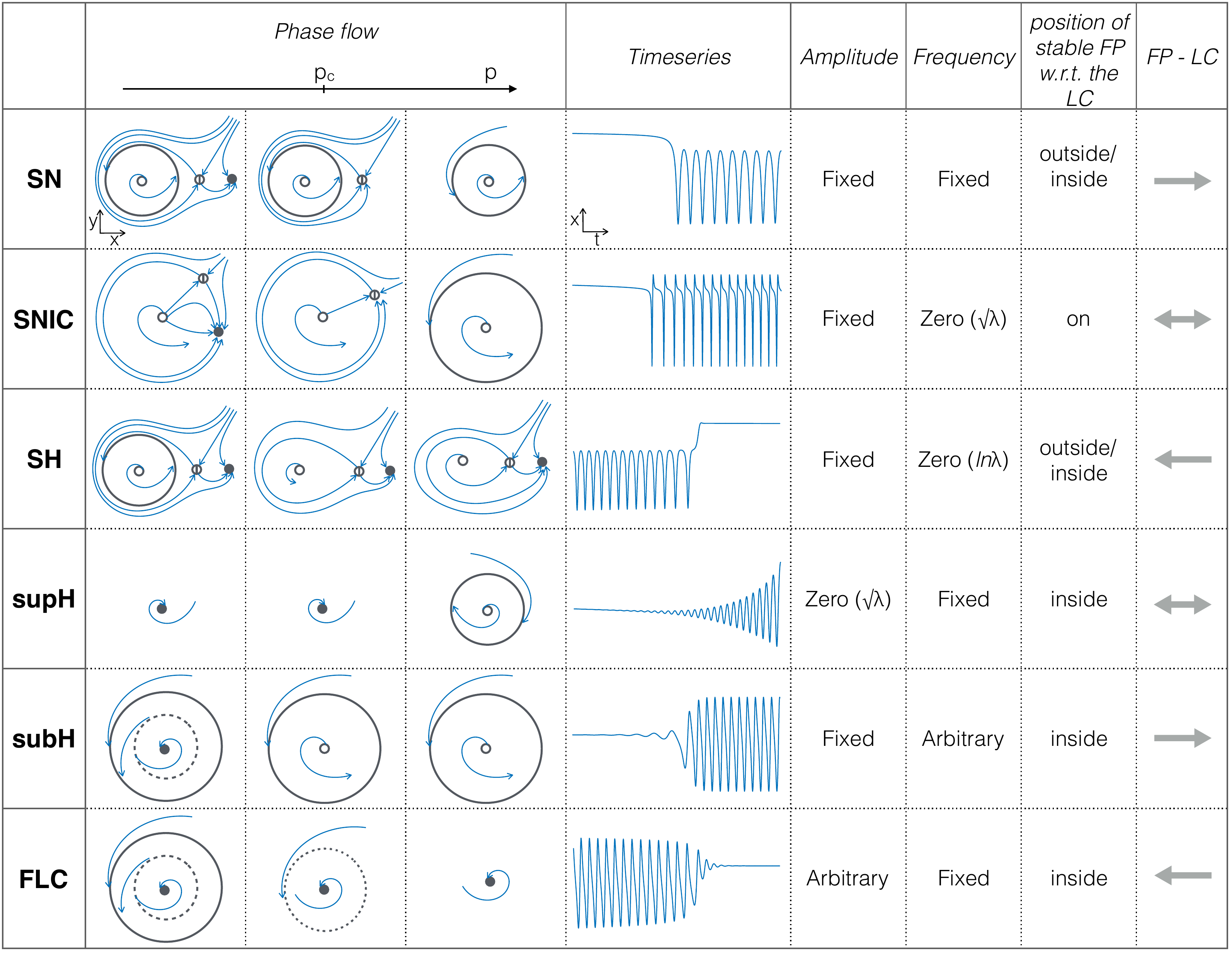} 
\caption{\textbf{Six planar bifurcations are responsible for oscillation onset and offset.} In a dynamical system, a bifurcation occurs when a smooth change of the values of some of the parameters of the system causes a sudden qualitative change of its behavior. The parameters that need to be varied to have this change in behavior are called bifurcation parameters, the number of bifurcations parameters necessary gives the \textit{codimension} of the bifurcation. In planar systems, only six types of codim-1 bifurcations can cause the transition from silent, that is a stable fixed point (FP) to oscillatory, that is a limit cycle (LC), behavior and/or viceversa. Their characteristics are illustrated in this figure. For each bifurcation we report an example of how the state space changes when varying the bifurcation parameter $p$. Bifurcations occur at the critical value $p_c$. Stable, saddle, unstable fixed points are represented by full, empty with a line inside, empty grey dots respectively. Stable, half-stable, unstable limit cycles are shown with solid, dotted, dashed lines respectively. Orbits appear in blue. We also show an example of timeseries and report the amplitude-frequency behavior, where $\lambda=p-p_c$. In the second last column we state whether the stable fixed point is inside or outside the stable limit cycle. This can affect the behavior of the baseline in the timeseries. The last column indicates the reversibility of the bifurcation, in the direction of the FP to the LC.}
\label{fig.6bif}
\end{figure}

Bursters come in different flavors. They can differ, among other factors, in the amplitude-frequency pattern of the active phase and in the behavior of the baseline. In the first formal classification of bursters, proposed by Rinzel \cite{rinzel1987formal}, the author used these features to determine the bifurcations responsible for oscillations onset and offset in the fast subsystem of bursters found in biological systems. 
This type of classification based on the onset/offset bifurcations pair has been later systematically extended by Izhikevich in his seminal paper \cite{izhikevich2000neural}, with the goal of including not only the known but also all the possible fast-slow bursters. His classification includes 120 different pairs of onset/offset bifurcations, of which 16 pertain to a planar fast subsystem with a fixed-point like silent state (resting-state could also be modeled with a small amplitude limit cycle). Izhikevich proposed to label each burster by stating the dimensionality of the fast and slow subsystem ($n+m$), the onset/offset bifurcations pair and whether the burster is of slow-wave or hysteresis-loop type. 

\begin{figure}[t]
\begin{minipage}{0.5\textwidth}
\includegraphics[width=\textwidth]{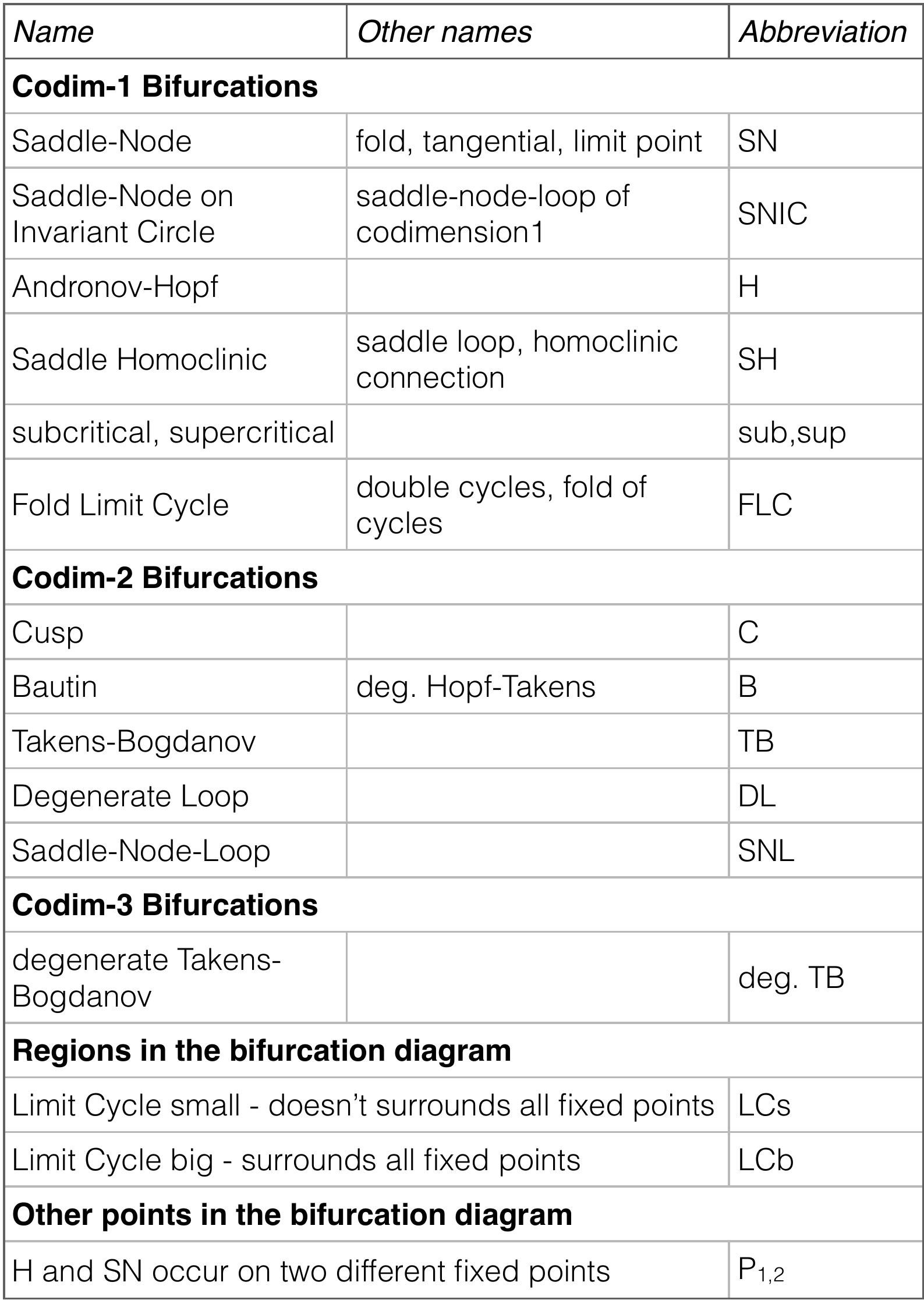}\vspace{0.5cm}
\end{minipage}\quad
\begin{minipage}{0.47\textwidth}
\includegraphics[width=\textwidth]{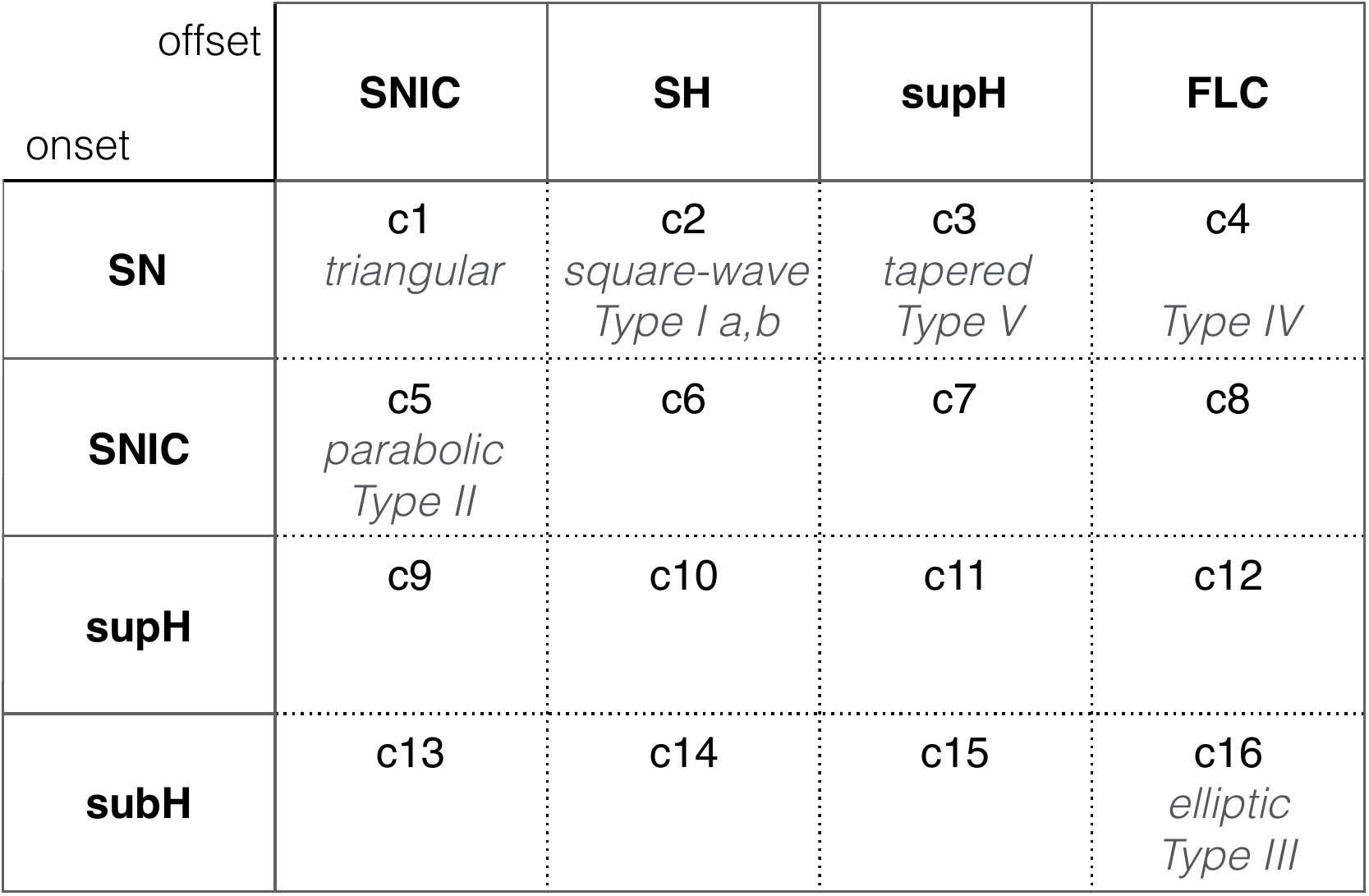}  
\caption{\textbf{Abbreviations.} In the left tabel we report the abbreviations used in this manuscript for bifurcations and regions in the bifurcation diagram. For the bifurcations we report alternative names available in the literature. The right table shows the sixteen classes for planar bursters identified by Izhikevich. Rows give the onset bifurcations, columns the offset bifurcations. We labeled each class with a `c' followed by a number from 1 to 16 as shown in the table. For each class we also report, when available, alternative names from the literature.}
\label{table_abbrev}
\end{minipage}
\end{figure}

In this work, we focus on bursters with the smallest dimensionality, namely $2+1$ for hysteresis-loop and $2+2$ for slow-wave. In both cases we have a planar ($n=2$) fast subsystem. In general, planar systems can exhibit only four codim-1 bifurcations (i.e. obtained by changing a single parameter) that allow the transition from a stable fixed point to a limit cycle, thus from the silent to the active phase. They are: saddle-node (SN), saddle-node-on-invariant-circle (SNIC), supercritical Hopf (supH) and subcritical Hopf (subH). Four bifurcations can be responsible for stopping the oscillation: SNIC, saddle-homoclinic (SH), supercritical Hopf and fold limit cycle (FLC). Considering all the pairs, we have sixteen different classes of planar bursters for slow-wave and sixteen for hysteresis-loop \cite{izhikevich2000neural}.

A description of the six planar bifurcations mentioned is given in Fig.~\ref{fig.6bif}. The tables in Fig.~\ref{table_abbrev} contain the abbreviations used in this paper. For brevity, we labeled the sixteen bursting classes by `c' followed by a number from 1 to 16, in the order of appearance in the table in Fig.~\ref{table_abbrev}. In the table we also report, when available, existing names for the classes.

\subsection{The unfolding theory approach}

The goal of the present work is to find a minimal descriptive model for bursters with a planar fast subsystem, for simplicity called planar bursters. We adopt a strategy developed by Golubitsky and colleagues \cite{golubitsky2001unfolding}, based on earlier work by Bertram et al. \cite{bertram1995topological} (see also de Vries \cite{de1998multiple}). 

Bertram and colleagues used as fast subsystem a model with a two-parameters bifurcation diagram, the Chay-Cook model for pancreatic $\beta$ cells bursting \cite{chay1988role}. They located, in this two-parameters bifurcation diagram, horizontal cuts crossing the codim-1 bifurcations curves required for some of the bursting classes known at that time. Horizontal cuts are straight paths in the parameter plane along which only one parameter is changing. This parameter is then used as slow variable. Using the same model, they could produce different classes by changing the location of the cut in the two-parameter bifurcation diagram.

This strategy has been later formalized by Golubitsky and colleagues \cite{golubitsky2001unfolding}. They realized that the codim-1 bifurcations of the fast subsystem which are necessary for bursting can be collapsed to a single local singularity of higher codimension, that is, a singularity in a high-dimensional parameter space, where the codim-1 bifurcations curves coincide. A path for bursting activity can then be found in the so-called unfolding of the singularity. 

The unfolding of a singularity of a dynamical system is a system that exhibits all possible bifurcations of that singularity \cite{Murdock:2006}. This unfolding can be described by adding some terms containing extra parameters to the normal form of the singularity. The number of extra parameters necessary, called \textit{unfolding parameters}, is the codimension of the singularity. In the unfolding parameter space there are manifolds (e.g. curves, surfaces) of lower codimension bifurcations points. These manifolds intersect at the origin, that is where all the extra parameters are zero and the system is equal to the normal form of the singularity. In the unfolding, we can search for paths that cross the right sequence of codim-1 bifurcations required by the burster, as done by Bertram and colleagues in the two-parameter bifurcation diagram of the Chay-Cook model. 

Let us consider the subH/FLC burster, for instance. For this class, no additional bifurcations, apart for those at onset and offset, are required to have hysteresis. We can thus take the unfolding of the codim-2 Bautin (also known as degenerate Hopf) singularity at which fold limit cycle and Hopf bifurcations occur together. In the unfolding, a curve of fold limit cycle bifurcations and a curve of Hopf (divided in a supercritical and a subcritical branch) stem from the Bautin point. We can thus locate a path for subH/FLC bursting. The path does not need to be horizontal, as long as it can be parametrized in terms of the slow variables. In this case, having hysteresis, one slow variable is enough.

The advantage of this approach is that we can use normal forms for the unfolding, if available, providing a minimal description for the fast subsystem. 

Golubitsky and coworkers systematically investigated the unfoldings of codim-1 and codim-2 bifurcations, with respect to bursting paths. They also extended the work to some regions close to a codim-3 singularity, but in a non-complete fashion. With regards to bursters with a planar fast subsystem, they identified nine slow-wave and three hysteresis-loop bursters. Hysteresis-loop can be harder to locate because, to exhibit hysteresis, they may require more bifurcations than their slow-wave counterpart. For example, consider the supH/supH burster, two supercritical Hopf bifurcations alone are not enough to create hysteresis, but the slow-wave burster can be built by going back and forth through a single supercritical Hopf point. On the other hand, hysteresis-loop bursters have a simpler mechanism, than slow-wave bursters, with regards to the slow dynamics. In slow-wave bursting the slow-subsystem must be at least two-dimensional and the path to follow in the ufolding must be completely specified. Hysteresis-loop bursting can be obtained with just one slow variable and is enough to specify the curve on which the path has to lie, while the points at which $z$ inverts its direction are determined by the crossing of the onset and offset bifurcation manifolds.

\subsection{Codim-3 degenerate Takens-Bogdanov singularity}

The codim-3 singularity used by Golubitsky et al. is called degenerate Takens-Bogdanov (deg. TB). Four topologically different unfoldings of this singularity have been identified by Dumortier et al. \cite{dumortier1987generic,dumortier1991bifurcations}. These unfoldings are very rich, containing saddle-node, SNIC, saddle-homoclinic, supercritical Hopf, subcritical Hopf and fold limit cycle bifurcations \cite{dumortier1991bifurcations}. This singularity had already appeared in the surroundings of models for neural bursting \cite{bertram1995topological,de1998multiple} and its biological importance has been further underlined by Osinga et al. \cite{osinga2012cross}. In one of the unfoldings of the deg. TB, the authors identified paths for many known bursters related to cell activity. They also implemented a slow-wave bursting model using a self-oscillating slow variable. 

In the present work we systematically extend Golubitsky and colleagues approach to the deg. TB singularity and investigate its four unfoldings. We aim at uncovering the presence or absence not only of paths for bursters known from cell activity, but of all planar bursters present in Izhikevich's classification. This would provide a general model ready to be applied in cell bursting and in any other fields, for which bursting classification is in progress, such as epileptic seizure modeling \cite{jirsa2014nature}. We give indications on how to build slow-wave bursters, which is in line with the work of \cite{osinga2012cross}. Furthermore we make use of hysteresis, when present, to build hysteresis-loop bursters. This allows less constraints on the required path and make it simpler to implement transitions between different bursting classes (see also Franci et al. \cite{franci2014modeling}).

A description of the planar codim-3 deg. TB singularity's equations and unfoldings has been provided by Dumortier and colleagues \cite{dumortier1987generic,dumortier1991bifurcations}. They identified four topologically different possibilities for this singularity and referred to them as codim-3 deg. TB cases: cusp, saddle, focus and elliptic. We investigated the three-parameters unfoldings of all four cases looking for possible paths for bursting activity, considering both the time-forward ($t\rightarrow\infty$) and time-reversal behaviors ($t\rightarrow-\infty$).

We found that the deg. TB singularity for the focus case in time-reversal condition gives the largest amount of bursting paths. Exploring the cusp, elliptic and saddle cases did not result in a description of new classes. For this reason, Section \ref{subsection_focus} is devoted to a detailed description of the focus case unfolding and its bursting paths. Later, in Section \ref{subsection_model}, we use this description to build a single model, which is able to display a vast repertoire of bursting activities.

Results for the cusp, saddle and elliptic cases are briefly summarized in Section \ref{subsection_othercases}.

\subsection{Unfolding the deg. TB singularity of focus case}
\label{subsection_focus}

\begin{figure}[t]
\includegraphics[width=\linewidth]{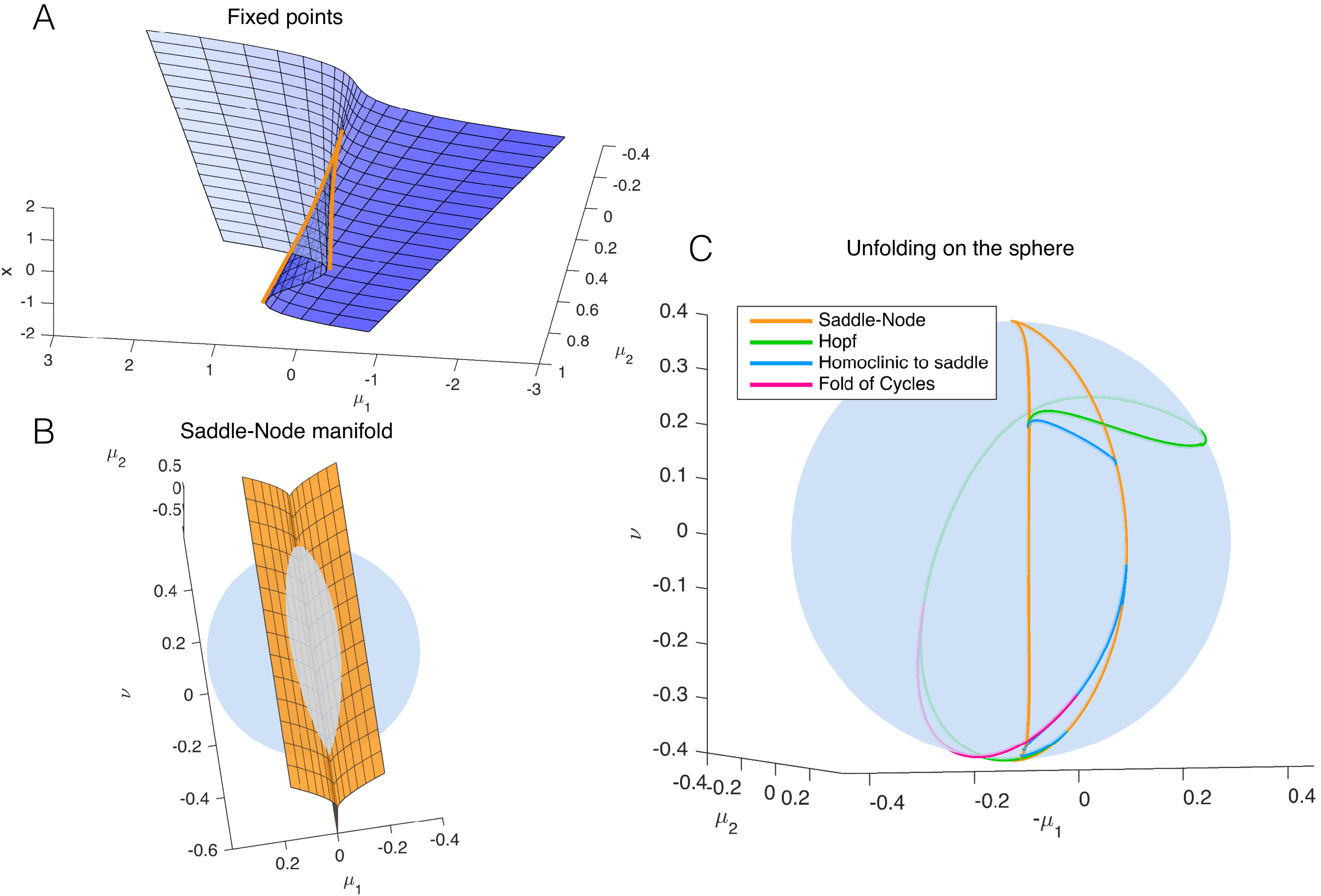} 
\caption{\textbf{Unfolding of the deg. TB singularity, focus case.} (A) The fixed points of the system are found for $y_0=0$, $x_0^3+\mu_2 x_0-\mu_1=0$. The blue surface represents $x_0$ plotted against the two parameters $\mu_1,\mu_2$. In orange are marked the curves of saddle-node bifurcations at which the saddle solution (i.e. the middle branch) collides with the focus in the upper or lower branch and annihilates. (B) The manifold of the  saddle-node bifurcation (in orange) is plotted in the three dimensional unfolding parameters space together with a sphere of radius $R = 0.4$ centered at the origin of the parameter space $(\mu_1,\mu_2,\nu)=(0,0,0)$. The intersection between the surface of  saddle-node bifurcations and the spherical surface gives a curve of saddle-node bifurcation (C) The bifurcation curves obtained at the intersection between the sphere and all the bifurcation surfaces of the unfolding.}
\label{fig_zeros_fold}
\end{figure}

The unfolding of the focus case in the time reversal condition is described by the following system of two coupled state variables $\left( x,y \right)$:

\begin{equation}
\label{eq_focus}
\begin{cases}
\dot{x}=-{y}\\
\dot{y}=x^3-\mu_2 x-\mu_1-y(\nu+b x+x^2)\\
\end{cases}
\end{equation}
where the dot above a variable describes the time derivative $d/dt$ and $\left( \mu_1, \mu_2,\nu \right)$ are the three unfolding parameters. The unfoldings obtained for any value of $b$ within the interval $0<b<2\sqrt{2}$ are topologically equivalent and correspond to the focus case \cite{dumortier1991bifurcations}. In the present work, we can thus set $b=1$ without any loss of generality. When $b>2\sqrt{2}$, instead, Eq.~\eqref{eq_focus} describe the elliptic case. Topological equivalence between the focus and elliptic cases has been shown by Baer and colleagues \cite{baer2006multiparametric}, which we will address in more details in Section \ref{subsection_othercases}.

We will explain here how moving in the unfolding parameter space $\left( \mu_1,\mu_2,\nu \right)$ affects the state space spanned by the variables $\left( x,y \right)$ \cite{dumortier1991bifurcations}.\\

\textbf{Representation on a sphere.} The codim-3 bifurcation occurs when the three unfolding parameters are equal to zero. Here saddle-node, Hopf, SNIC, saddle-homoclinic and fold limit cycle bifurcations coincide. From the origin of the parameter space, surfaces for codim-1 bifurcations arise. At the intersection between surfaces of codim-1 bifurcations we have curves of codim-2 bifurcations.

\begin{figure}[t]
\centering
\includegraphics[width=0.65\textwidth]{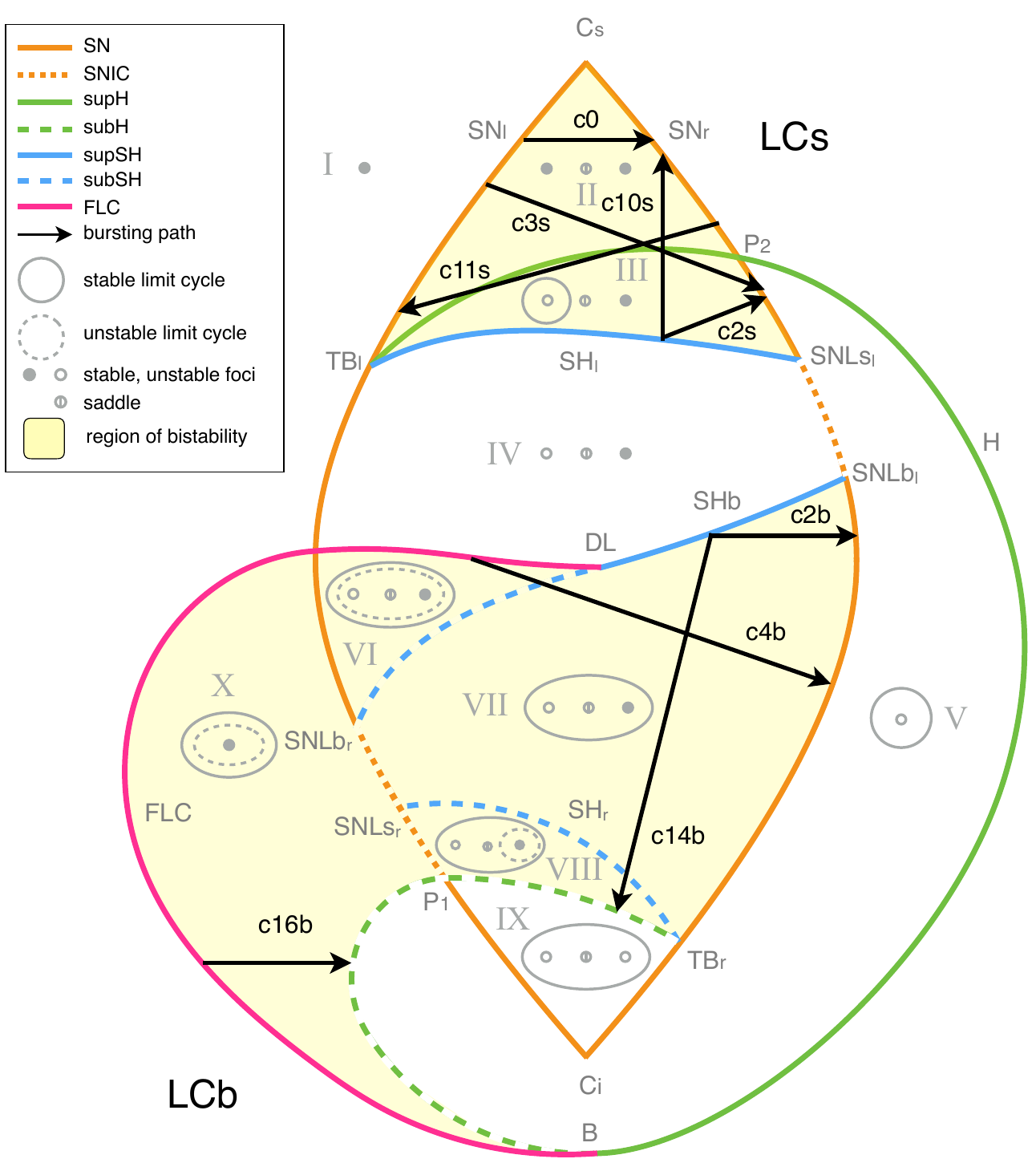} 
\caption{\textbf{Paths for hysteresis-loop bursting activity in the unfolding of the deg. TB singularity of focus type.} This is a flat representation topologically equivalent to the bifurcation diagram shown in Fig.~\ref{fig_zeros_fold}C. The five bifurcation types, saddle-node, SNIC, Hopf, saddle-homoclinic and fold limit cycle are depicted in orange, dashed orange, green, light blue, and dark blue, respectively. For each region of the diagram, labeled with a Roman numerals from I to X, a schematic description of the phase portrait is offered in grey, using solid/dashed circles to represent stable/unstable limit cycle, full/empty dots for stable/unstable foci and an empty dot with a line in the middle to represent saddles. A more detailed description of these ten regions is given in Fig.~\ref{fig_flows}. There are two separate regions of bistability (in yellow), one in the lower part of the unfolding, where the stable limit cycle is big enough to surround all the fixed points existing, labeled LCb (limit cycle big) region; the other in the upper part, where the limit cycle does not surround all the fixed points, labeled LCs region.  Paths for bursting activity are drawn as black arrows. The direction is chosen so that the path encounters first the offset and then the onset bifurcations. When building the model, this will be the direction along which the slow variable increases.}
\label{fig_focus_bifdiagr}
\end{figure}

\begin{figure}[t]
\center
\includegraphics[width=0.72\textwidth]{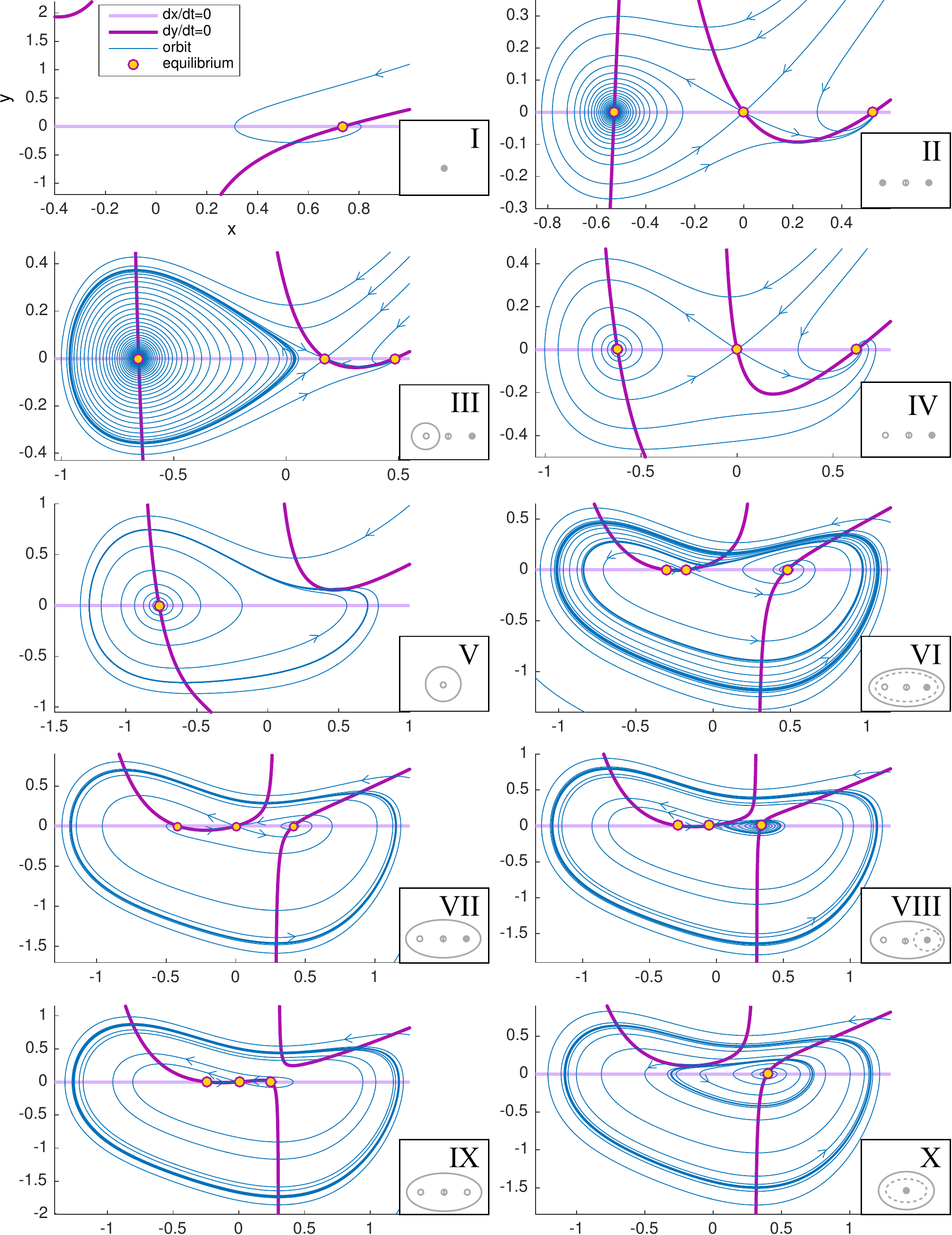}
\caption{\textbf{Phase flows.} For each region of the unfolding, labeled with a Roman numeral in Fig.~\ref{fig_focus_bifdiagr}, we show the nullclines of Eq.~\eqref{eq_focus} in light and dark purple. Fixed points at the intersection of the nullclines are marked with yellow dots. Flows are shown in blue. Nullclines and fixed points are computed analytically, flows are obtained through numerical simulations in Matcont.}
\label{fig_flows}
\end{figure}

To describe what happens in the surrounding of the codim-3 singularity we can consider the intersection of the bifurcation curves and surfaces with a sphere centered around the singularity at the origin. On the sphere we have curves of codim-1, points of codim-2 and no codim-3 bifurcations. The bifurcation portrait on the sphere is topologically equivalent for any sufficiently small value of the radius $R$ \cite{dumortier1991bifurcations}. Using a fixed radius allows for a description of the bifurcations with two parameters, that is, the spherical $\left(
 \theta,\phi \right)$ spherical coordinates, instead of three parameters $(\mu_1,\mu_2,\nu)$ in cartesian coordinates. The result of the numerical evaluation of the unfolding on the sphere (obtained using Matcont and CL Matcont for numerical continuation of codim-1 bifurcations) is shown in Fig.~\ref{fig_zeros_fold}C, and an example for a, topologically equivalent, cartoon flat representation in Fig.~\ref{fig_focus_bifdiagr}. For each region of the unfolding, labeled with Roman numerals in Fig.~\ref{fig_focus_bifdiagr}, we computed the nullclines and perfomed simulations for different initial conditions. These results are shown in Fig.~\ref{fig_flows}.\\

\textbf{Fixed points and local bifurcations.} The system is in a fixed point, or equilibrium, $(x_0, y_0)$ when invariant with respect to time t, that is $\dot{x}=0,\dot{y}=0$. The corresponding solution for Eq.~\eqref{eq_focus} is $y_0=0, x_0^3+\mu_2 x_0-\mu_1=0$. Hence, it appears that the fixed points do not depend on $\nu$. $x_0$ is displayed in Fig.~\ref{fig_zeros_fold}A as a function of ($\mu_1$,$\mu_2$). We can distinguish two regions in the space $(\mu_1\,\mu_2)$: one in which a single fixed point, a focus, exists; the other in which we have a focus on an upper branch, a focus on a lower branch and a saddle in a middle branch. The saddle coalesces with the focus of the upper branch along the saddle-node bifurcation curve $SN_r$ and with the focus of the lower branch along $SN_l$. Right and left (and later inferior, superior) refer to where the bifurcation occurs in the state space. Fig.~\ref{fig_zeros_fold}B shows the saddle-node bifurcation in the complete parameter space of the unfolding and its intersection with the sphere: a closed curve which delimitates the region with three fixed points from the region with a single fixed point. Two codim-2 cusp bifurcations, $C_s$ and $C_i$, occur in the place where $SN_l$ and $SN_r$ meet and vanish.

The condition for the Hopf bifurcation can be found by equating the trace of the Jacobian (at the fixed point) to zero. The Hopf bifurcation takes place if  $x^2+x+\nu=0$ and $x^3-\mu_2 x-\mu_1=0$. The Hopf bifurcation in the unfolding is represented by green lines in Fig.~\ref{fig_zeros_fold}C and Fig.~\ref{fig_focus_bifdiagr}, where a solid/dashed line is used for the supercritical/subcritical cases. On the sphere, we have two codim-2 Takens-Bogdanov bifurcation points where the Hopf and saddle-node bifurcations, $TB_l$ on $SN_l$ and $TB_r$ on $SN_r$, meet. Note that the other two intersections between the Hopf and saddle-node bifurcation curves in Fig.~\ref{fig_focus_bifdiagr}, $P_1$ and $P_2$, are not Takens-Bogdanov points as the two bifurcations act on two different foci.\\

\textbf{Global bifurcations.} Results for the global bifurcations are obtained numerically. A single stable limit cycle exists in the system given by Eq.~\eqref{eq_focus}. To describe the unfolding, we can consider the stable limit cycle originating at the supH curve, between $TB_l$ and the Bautin point $B$, and we can follow its evolution and annihilation. Starting at the $TB_l$, the limit cycle arises from the destabilization of the stable focus on the lower branch and grows until it meets the saddle in the middle branch. Here the limit cycle vanishes and we have a curve of saddle-homoclinic bifurcations $SH_l$, which starts at $TB_l$ and terminates on $SN_r$ giving rise to the codim-2 saddle-node-loop bifurcation $SNLs_l$. `s' denotes a small limit cycle, in the sense that it does not surround all the fixed points. From $SNLs_l$ to $SNLb_l$, the limit cycle disappears through a SNIC bifurcation giving rise to a heteroclinic channel between the saddle and the stable focus appeared through $SN_r$. $SNLb_l$ marks the point where the limit cycle has grown big enough to encircle all the fixed points. From here to the  $DL$ (degenerate loop) point, in fact, the limit cycle disappears through a `big saddle-homoclinic' bifurcation $SHb$ (the saddle-homoclinic bifurcation is said `big' if the limit cycle encompasses the stable fixed point \cite{kuznetsov2013elements,izhikevich2000neural}). After $DL$, the limit cycle is not able to reach the saddle anymore and coalesces with an unstable limit cycle on the fold limit cycles curve $FLC$. This unstable limit cycle, which is always enclosed by the stable one, can originate in two ways: from the subcritical branch of the Hopf curve or from the subcritical branch of $SHb$. The unstable limit cycle can also disappear before reaching the $FLC$ curve, via a SNIC bifurcation, from $SNL_r^{s}$ to $SNL_r^{b}$.

\section{Results}
\label{sec_results}

\subsection{Hysteresis-loop bursting classes}

We investigated the two-parameters bifurcation topology (i.e. the unfolding on the sphere) to identify paths for bursting activity. We propose to consider the system given in Eq.~\eqref{eq_focus} as the fast subsystem, which is moved by a slow subsystem in the parameters space so that bursting occurs.

In the present work we are particularly interested in bursters driven by a single slow variable, which oscillates due to feedback from the fast subsystem. For this purpose, the state space of the fast subsystem must display hysteresis between the silent and the active states. The slow variable can be instructed, in the simplest form by linear feedback, to steer the path in a given direction when the system is close to a stable fixed point representing the silent phase, and in the opposite direction when the system has moved to an attractor far from the silent phase. If this second attractor is a limit cycle, the system is in the active phase.

A prerequisite of hysteresis is the existence of a regime in which at least two stable states coexist, that is, bistability. We find two regions on the sphere where bistability occurs (in yellow in Fig.~\ref{fig_focus_bifdiagr}). One region in the lower portion of the bifurcation diagram, where the limit cycle surrounds all the fixed points, which we named LCb (limit cycle big) region. The other region is in the upper part of \ref{fig_focus_bifdiagr}, here the limit cycle does not surround all fixed points. We named it LCs (limit cycle small) region. We added `b' or `s' to the label the regions where bifurcations and bursting classes occur.\\

\begin{figure}
\center
\includegraphics[width=0.95\textwidth]{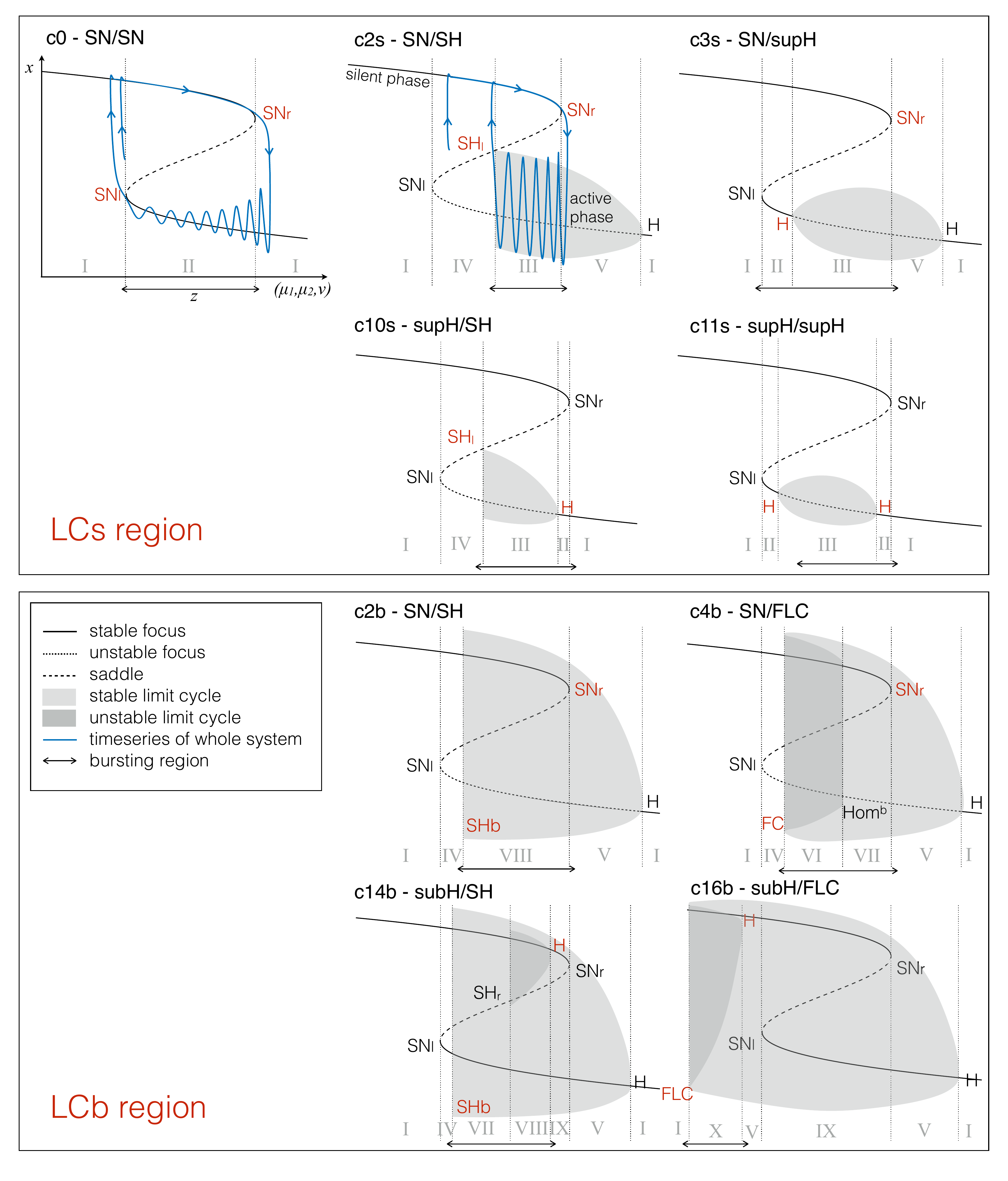} 
\caption{\textbf{Bifurcation diagrams along bursting paths.} For each of the black bursting paths in Fig.~\ref{fig_focus_bifdiagr} we propose a cartoon representation of the bifurcation diagram of the fast subsystem for the fast variable $x$ and using $z$ as bifurcation parameter. Onset and offset bifucations are written in red. In all the nine classes (eight point-cycle and one point-point bursters) the upper branch of the z-shaped curve of fixed points acts as silent state. When the fast subsystem is in the silent state, the slow variable $z$ is instructed to increase.  At $SN_r$ (or at the subcritical Hopf point in classes c14b and c16b) the resting-state destabilizes, the system moves towards another attractor, and $z$ start decreasing until the system goes back to the silent state and a new bursting cycle is started. The active phase takes place on the limit cycle generated by a supercritical Hopf bifurcation on the lower branch. The top panel shows bifurcation diagrams for bursters in the LCs region, the bottom panel for those in the LCb region. In the LCs region we can have a saddle-node onset if the limit cycle exists already when the silent state destabilizes at $SN_r$ (first row) or supercritical Hopf onset when the limit cycle is created for smaller values of $z$ (second row). Oscillations can end because the limit cycle coalesces with the saddle (middle column) or because of another supercritical Hopf bifurcation (right column). In the LCb region panel, in the first row oscillations are started through saddle-node bifurcation, while the resting-state destabilizes earlier in the second row via subcritical Hopf bifurcation. The limit cycle coalesces with the saddle in the left column and with an unstable limit cycle in the right column. }
\label{fig_classes}
\end{figure}

\textbf{LCs bursters.} In the region LCs, oscillations can start through the SN bifurcation ($SN_r$ between $SNL_l^{s}$ and $P_2$) or the supH. The limit cycle can vanish through the supercritical Hopf or the saddle-homoclinic bifurcations. Consequently, we considered and verified the existence of four pairs of onset/offset bifurcations: c2s (SN/SH), c3s (SN/supH), c10s (supH/SH) and c11s (supH/supH). This region contains in addition to this four cases a special case of burster in which no limit cycle exists and both active and silent phases are given by fixed points (point-point burster \cite{izhikevich2000neural}). In this case both onset and offset are given by the saddle-node bifurcation. When the stable focus, which represents the silent phase, destabilizes, the system spirals towards the other stable focus. This spiraling is the active phase. We attributed the number 0 to the SN/SN bursting class, which is not among the sixteen point-cycle classes. Typical paths are indicated by black arrows in Fig.~\ref{fig_focus_bifdiagr} and examples of the bifurcation diagrams are in the top panel of Fig.~\ref{fig_classes}.\\

\textbf{LCb bursters.} In the LCb region oscillations can be generated by the saddle-node bifurcation ($SN_r$) or the subcritical Hopf. Oscillations can be stopped through the saddle-homoclinic or the fold limit cycle bifurcations. Consequently, we considered and verified the existence of four pairs of onset/offset bifurcations: c2b (SN/SH), c4b (SN/FLC), c14b (subH/SH) and c16b (subH/FLC). 
Typical paths are shown in Fig.~\ref{fig_focus_bifdiagr}, and examples of the bifurcation diagrams are shown in the bottom panel of Fig.~\ref{fig_classes}. 

\subsection{Hysteresis-loop bursters: a unique model }
\label{subsection_model}

\begin{figure}
\center
\includegraphics[width=\textwidth]{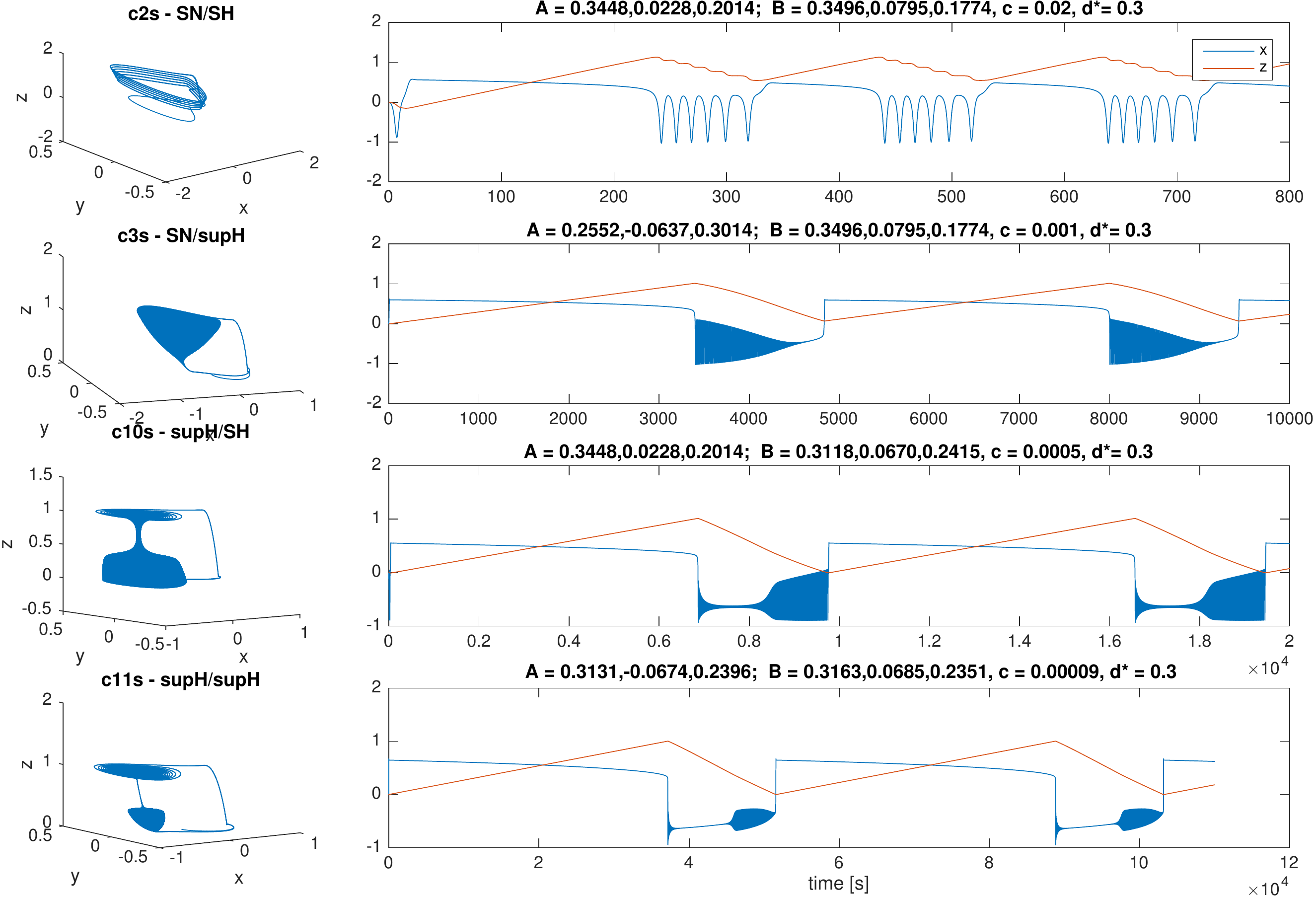}
\includegraphics[width=\textwidth]{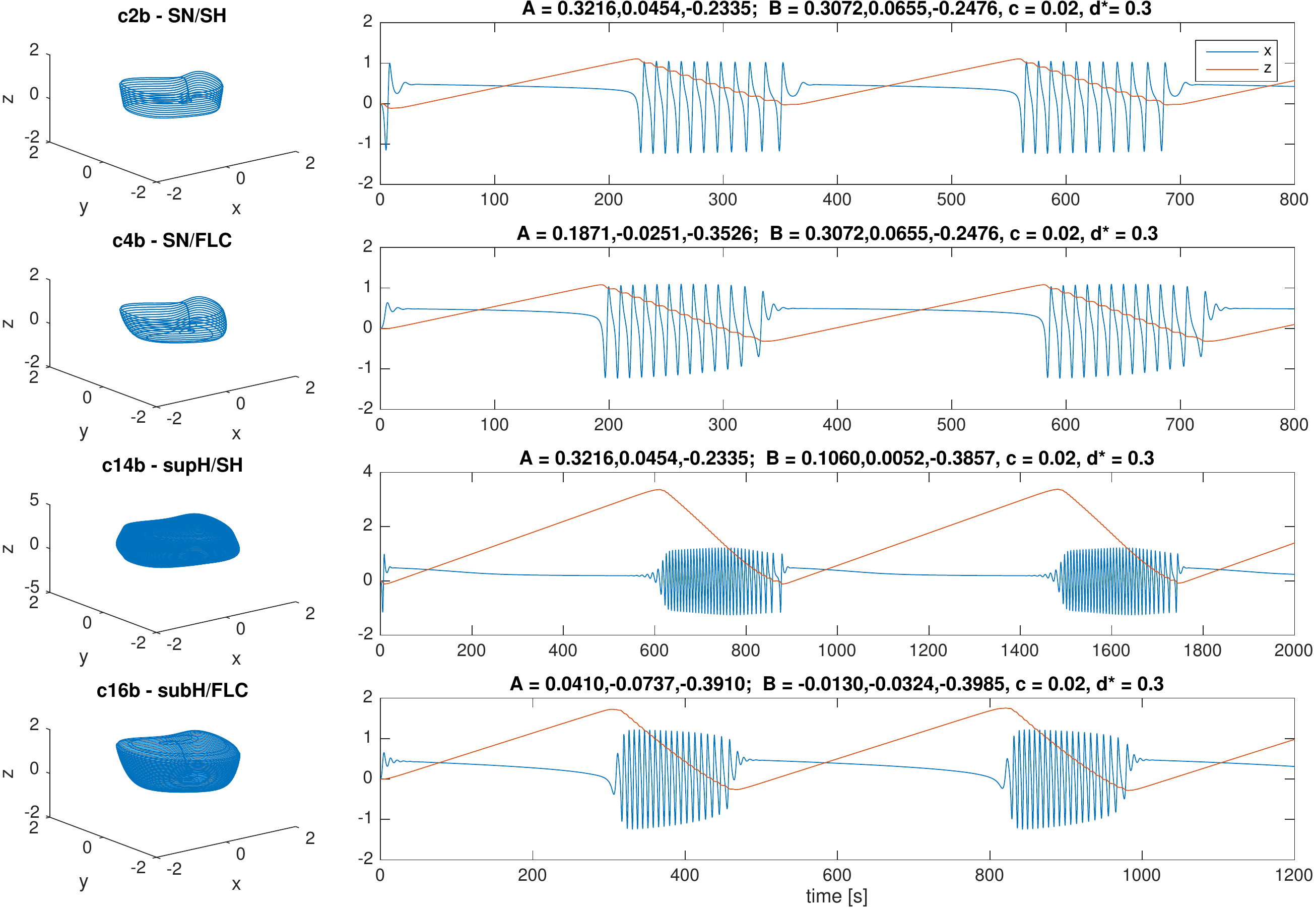} 
\caption{\textbf{Different classes.} For each of the classes identified is portrayed the evolution in the phase space (right panel) and the timeseries for $x$ and $z$ (right panel). To obtain different classes we tuned the parameters A and B which affect the position of the path in the unfolding.}
\label{fig_simulations}
\end{figure}

The canonical form of the unfolding of the codim-3 deg. TB singularity, focus case, time reversal, is

\begin{equation}
\begin{cases}
\dot{x}=-{y}\\
\dot{y}=x^3-\mu_2 x-\mu_1-y(\nu+x+x^2)\\
\end{cases}
\end{equation}

For bursting activity we need the fast subsystem to slowly move in the unfolding parameter space following a path to undergo the required bifurcations. We can parametrize this path in terms of a third variable $z$, which slowly changes in time. This variable steers the system through the parameter space and drives it into and out of oscillatory behavior.
In particular, when the distance in phase space between the state of the system $( x,y )$ and the silent state $(x_{s}(z),0)$ is smaller than a certain threshold $d^*$ the system should move to the point in parameter space where the silent state loses stability. On the other hand, when the distance between the state of the system and the silent state is bigger than $d^*$ the system should move to the point in the parameter space where the limit cycle destabilizes. In other words, when the system is silent, it has to move in the direction of the bursting onset bifurcation, when it is active, it has to move towards the bursting offset bifurcation. If we consider a curve in the parameter space, which starts at the offset bifurcation and extends towards the onset bifurcation (as indicated by the black arrows in the bifurcation diagrams in Fig.~\ref{fig_focus_bifdiagr}), $\dot{z}$ should be positive when the system is in the silent state ($\sqrt{(x-x_{s}(z))^2+y^2}<d^*$) and positive otherwise. The dynamics of $z$ can thus be described by:

\begin{equation}
\begin{cases}
\dot{x}=-{y}\\
\dot{y}=x^3-\mu_2(z) x-\mu_1(z)-y(\nu(z)+x+x^2)\\
\dot{z}=-c(\sqrt{(x-x_{s}(z))^2+y^2}-d^*)
\end{cases}
\end{equation}

where $c$ is the velocity at which $z$ changes along the path. 

As described in Section \ref{subsection_focus}, the unfolding parameters can be reduced to two if we restrict the movements to a spherical surface centered at the codim-3 singularity. We can perform this reduction without loss of generality because the bifurcations curves on the sphere will be topologically equivalent to those of any other sphere, providing a small enough radius. 

With the coordinates transform for the unfolding parameters described by

\begin{equation}
\begin{cases}
\mu_2 =  R\sin{\theta}\cos{\phi}\\
\mu_1 = -R\sin{\theta}\sin{\phi}\\
\nu=R \cos{\theta}
\end{cases}
\end{equation}

the model reads

\begin{equation}
\label{model_eqs}
\begin{cases}
\dot{x}=-{y}\\
\dot{y}=x^3-R \sin{\theta(z)}\cos{\phi(z)} x+R\sin{\theta(z)}\sin{\phi(z)}-y(R\cos{\theta(z)}+x+x^2)\\
\dot{z}=-c(\sqrt{(x-x_{s}(z))^2+y^2}-d^*)
\end{cases}
\end{equation}

The simplest curve satisfying the requirements, considering that our 2-dimensional parameter space lies on a spherical surface, is the shortest arc on this surface between the initial and final point, known as great circle.

\begin{figure}
\center
\includegraphics[width=\textwidth]{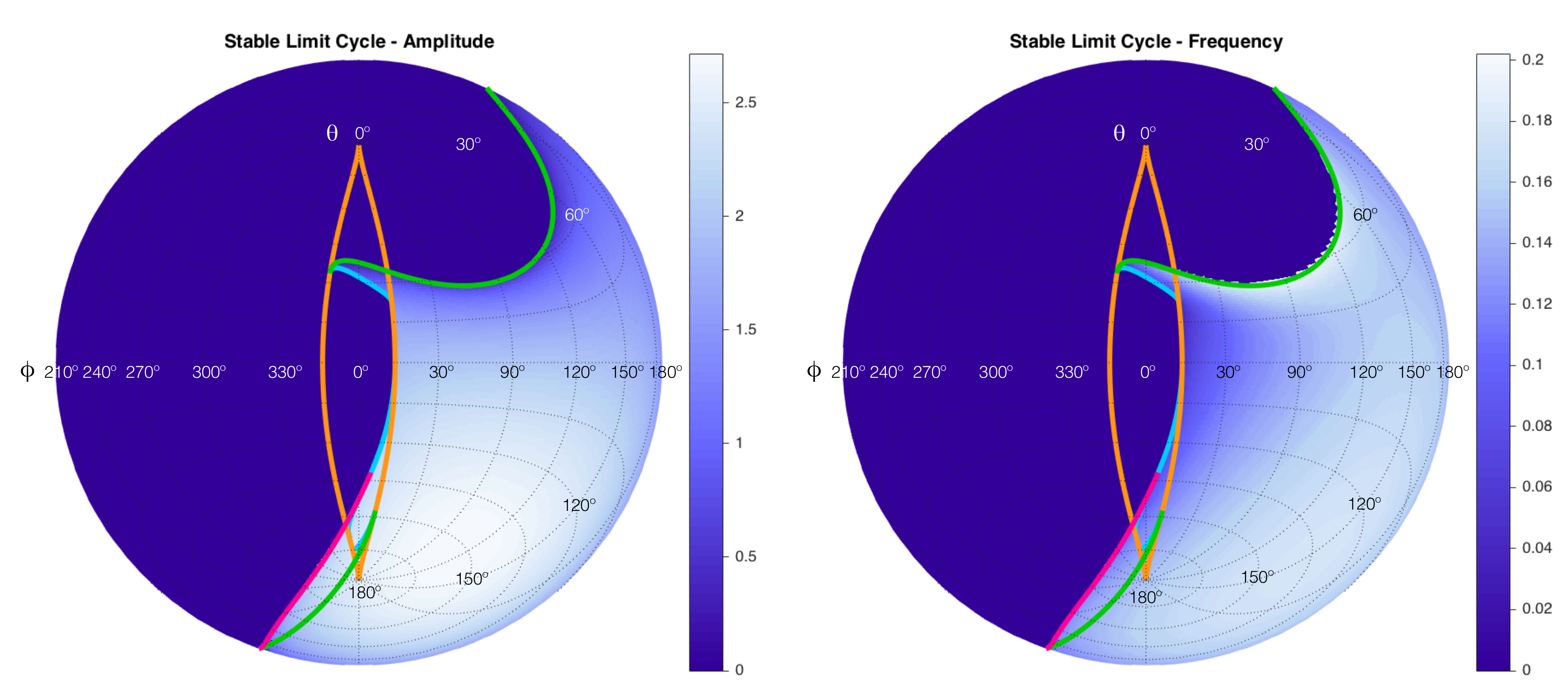}
\caption{\textbf{Stable limit cycle of the focus case.} We investigated the amplitude (left panel) and frequency (right panel) of the single stable limit cycle present in the focus case. To obtain a flat representation of the spherical surface we here make use of Lambert Equal Area Azimuthal Projection.}
\label{fig_statespace}
\end{figure}

To provide a parametrization of the great circle we consider the parametric equations of the line through points $A$ and $B$ in cartesian coordinates $( \mu_2,-\mu_1,\nu )$:

\begin{equation}
\label{eq_line}
\begin{cases}
\mu_2(z)=\mu_{2,A}+(\mu_{2,B}-\mu_{2,A})z\\
-\mu_1(z)=\-\mu_{1,A}-(\mu_{1,B}-\mu_{1,A})z\\
\nu(z)=\nu_{A}+(\nu_{B}-\nu_{A})z
\end{cases}
\end{equation}

Points on this line and on the corresponding great circle have the same angles $\theta$ and $\phi$ but a different radius, fixed and equal to $R$ in the great circle case. The parametric equations of the corresponding great circle in spherical coordinates are thus

\begin{equation}
\label{eq_arc}
\begin{cases}
r(z)=R\\
\theta(z)=\arccos\left(\dfrac{\nu(z)}{R}\right)\\
\phi(z)=\arctan\left(\dfrac{-\mu_1(z)}{\mu_2(z)}\right)\\
\end{cases}
\end{equation}

This parametrization (Eq.~\eqref{eq_line}-\ref{eq_arc}) and Eq.~\eqref{model_eqs} provide a model able to reproduce all hysteresis-loop bursting classes found in the unfolding of the deg. TB singularity, focus case.

\begin{figure}[t]
\center
\includegraphics[width=0.96\textwidth]{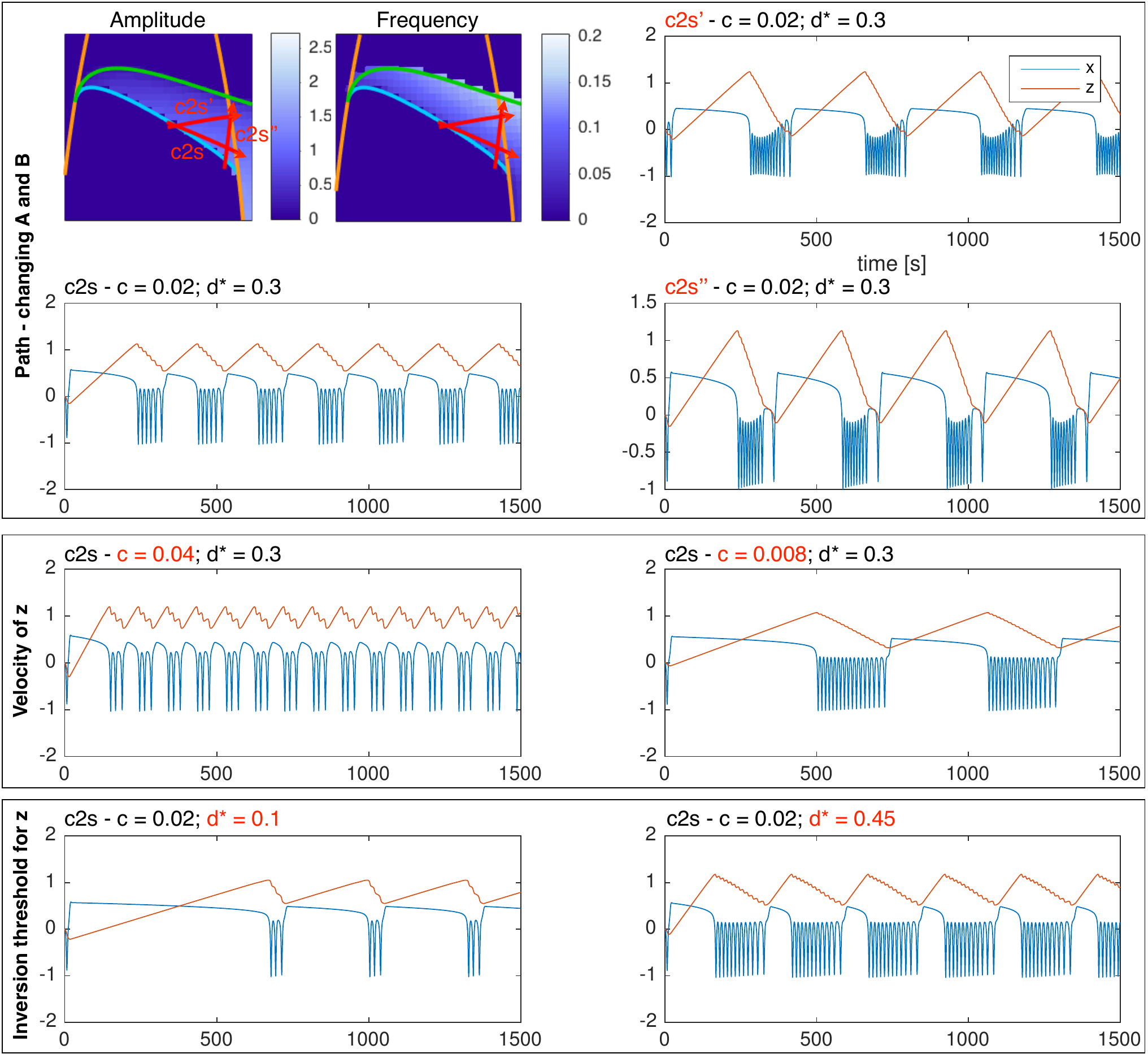}
\caption{\textbf{Effect of the parameters of the model on the timeseries of a class.} We show how different realizations of the same class can be obtained by changing the parameters of the model. We use as example c2s (SN/SH). In the top panel we show how the amplitude-frequency profile of a given class can be changed by choosing different paths (c2s, c2s', c2s'') in the unfolding. The only requirement is that the starting point A of the path has to lay on the $SH_l$ curve and the ending point B on $SN_r$ curve between $SNLs_l$ and $p_1$. In the middle panel we show how the velocity of $z$, $c$ affects the length of both the active and silent phases. In the bottom panel we can see how changing $d^\star$ affects the active/silent phases ratio. The values for A and B used to obtain the simulations are: for c2s A=[0.3448,0.02285,0.2014], B=[0.3496,0.07955,0.1774];  for c2s' A=[0.3448,0.02285,0.2014], B=[0.3331,0.074,0.2087]; for c2s'' A=[0.3551,0.07019,0.1703], B=[0.3331,0.074,0.2087].}
\label{fig_effectparam}
\end{figure}

To summarize the elements appearing in the model, we have:

\begin{itemize}
\item A fast subsystem $( x,y)$, which in some regions of the unfolding space presents bistability and hysteresis between a stable fixed point and a stable limit cycle.
\item A slow subsystem $z$, which depends on the feedback from the fast subsystem and moves the latter through the parameter space. When the fast subsystem shows hysteresis, the slow one can drive it in and out from the oscillatory behavior giving rise to bursting activity.
\item The constant $c$ determines the speed of the movement of the subsystem in the parameter space, as promoted by $z$. It should be small enough to guarantee timescale separation between the fast and slow subsystems. This constant affects the length of both silent and active phases, and thus the number of oscillations in the active phase.
\item $x_{s}(\theta(z),\phi(z))$ is the $x$ coordinate of the upper branch equilibrium point of the fast subsystem, which acts as the silent state.
\item $d^*$ is the threshold that determines the distance of the fast subsystem from the silent state required to promote a change in the direction of $z$. It should be smaller than the shortest distance between the silent state and the limit cycle. The smaller it is, the bigger is the silent/active phase length ratio. 
\item $R$ is the radius of the shell of the unfolding around the codim-3 singularity. All results in this work are for $R=0.4$.
\item $\theta$ and $\phi$ are the solid angle in the spherical coordinates of the unfolding. They have been parametrized in terms of $z$ to follow an arc of great circle determined by two points $A$ and $B$ on the sphere.
\item $(\mu_2$,$-\mu_1$,$\nu)$ are the cartesian coordinates in the unfolding and should be provided for the point $A$ of offset bifurcation and for that $B$ of onset bifurcation. These are the values that determine the bursting class the system belongs to, if any. The points A and B are used to determine the great circle on which the arc has to lie, the direction of movement (from A to B) and the initial point A. But the actual final point of the path is automatically given from the mechanisms that force $z$ to change direction and is thus close to the onset bifurcation.
\end{itemize}

We performed simulations for each of the classes found in the focus case, as shown in Fig.~\ref{fig_simulations}. For each class, the evolution in the phase space is also shown.

The same class can be obtained in different ways, by changing the location of the A and B points on the offset/onset bifurcation curves respectively. To explore the effects of the location of the path on the shape of the timeseries, we investigated how the amplitude and frequency of the stable limit cycle change across the unfolding. Results are shown in \ref{fig_statespace}. By choosing different paths it is possible to alter the amplitude-frequency profile of timeseries, within the constraints imposed by a given class.

As an example, we show in Fig.~\ref{fig_effectparam} three different realizations for the class c2s. In the same figure we also show the impact of varying the velocity, $c$, at which $z$ moves along the path and the effect of different choices of the parameter $d^*$, which determines when $z$ inverts direction.

\subsection{Transitions between classes}

\begin{figure}[t]
\includegraphics[width=\textwidth]{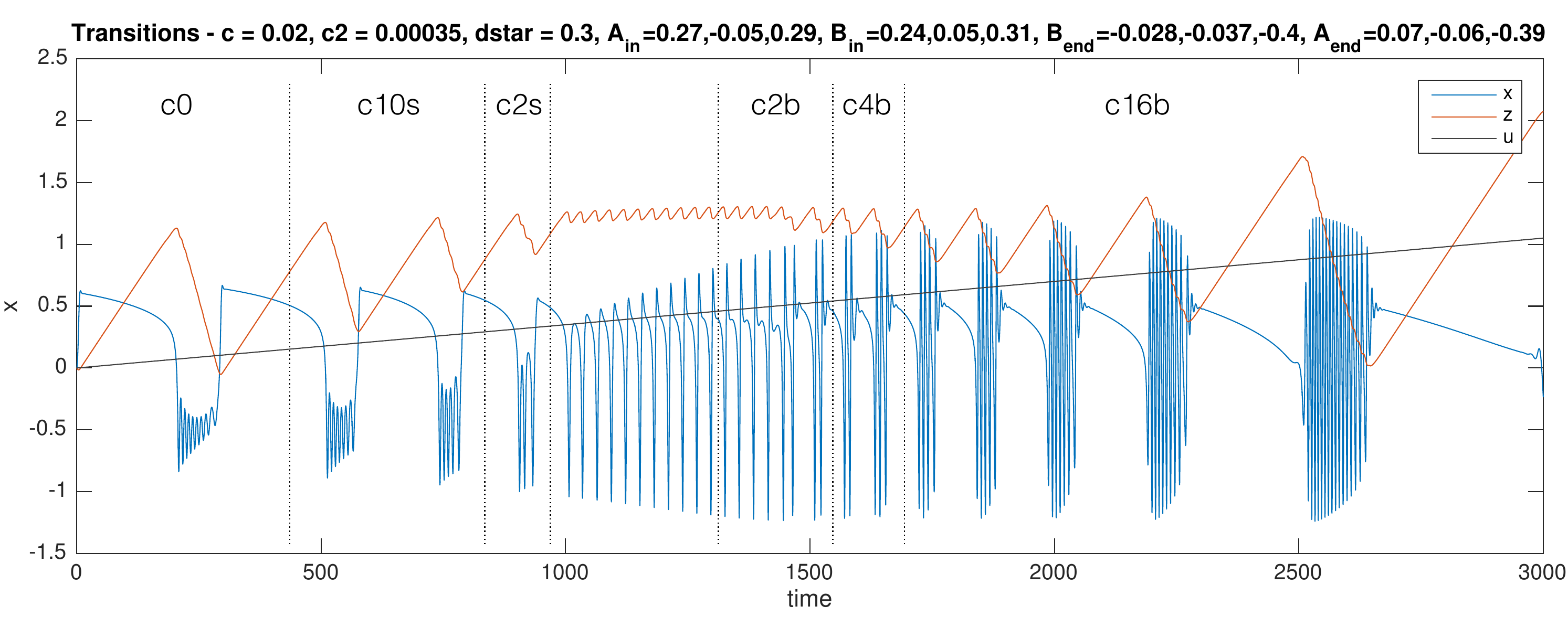} 
\begin{minipage}{0.5\textwidth}
\includegraphics[width=\textwidth]{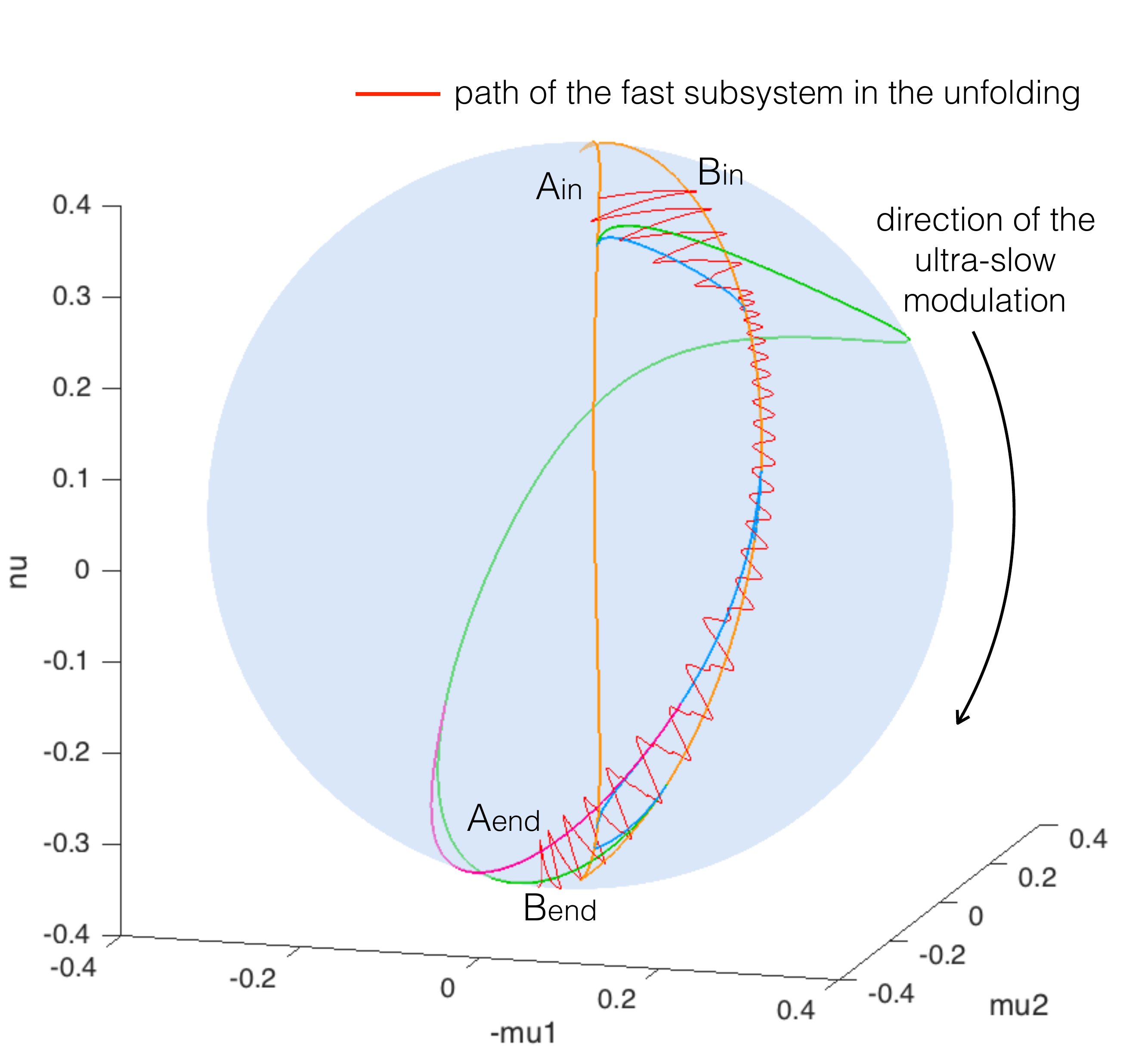} 
\end{minipage}
\begin{minipage}{0.5\textwidth}
\caption{\textbf{Transitions among classes.} Transitions between bursting classes can be obtained through an ultra-slow modulation of the starting and ending points of the path followed by the fast subsystem in the unfolding. In the top panel is shown the simulated timeseries for the fast variable $x$, the slow one $z$ and the ultra-slow modulation $u$. Dotted vertical lines mark the separation between different bursting classes. The path followed in the unfolding is displayed in red in the bottom panel. In the example portrayed, bursting activity moves from the LCs region (SN/SN, supH/SH, SN/SH) in the upper part of the unfolding, through a region with only oscillatory behavior, to the LCb region (SN/SH, SN/FLC, subH/FLC) in the lower part of the unfolding.}
\label{fig_transitions}
\end{minipage}
\end{figure}

The fact that the hysteresis-loop model produces bursting activity of one class rather than another only depends on the path followed in the unfolding, in our case, on the arc on the sphere. The settings that, in the present work, determine this arc are the two points A and B. Ultra-slow modulations of these points can determine a change of classes. An example is shown in Fig.~\ref{fig_transitions}, where we have implemented a straight downward ultra-slow movement of the initial and final points of the path. The system is initially in the LCs region in the upper part of the unfolding. The initial bursting class is c0, changes to c3s, then c2s to end in a region with just oscillatory, not bursting, activity. Note that in this region, by construction, timescale separation does not hold any more. The system then enters the LCb region in the bottom part of the unfolding starting with c2b, through c4b to c16b.

In general, transitions of classes are possible within the same region (LCs or LCb). To have a transition between classes in different regions, the system has to go through a simple oscillatory phase or a simple silent phase (as for example in the central part of Fig.~\ref{fig_transitions}).

It is important to stress that $A$ and $B$ are not the ending points of the path followed by the system, but determine the great circle it lies on, the direction of movement and the starting point. As $z$ inverts its direction after onset and offset bifurcation, those bifurcation points have the role of limiting the arc followed by the system. For this reason, even though in Fig.~\ref{fig_transitions} we moved $A$ and $B$ downwards following the arcs connecting $A_{in}$ to $A_{end}$, and $B_{in}$ to $B_{end}$, respectively, the actual zig-zag path followed by the fast subsystem is determined by the onset and offset bifurcations it undergoes.

This is not true for slow-wave bursters, in which the whole trajectory in the unfolding must be specified. For this reason, and for the fact that path shapes are specific to each class, transitions between these kind of bursters can be more difficult to implement.

\subsection{Slow-wave bursters}

\subsubsection{Slow-wave bursting classes}

Slow-wave classes do not require hysteresis and are driven by at least two slow-variables, which makes it possible to use closed paths along which the system will move in a given direction.  Examples of paths found for the slow-wave classes are shown in Fig.~\ref{fig_focus_bifdiagr_sw}. When paths for hysteresis-loop bursting exist, they can be used to implement slow-wave bursters as well. For this reason, we show in Fig.~\ref{fig_focus_bifdiagr_sw} only the classes that do not have a hysteresis-loop counterpart in our model.

\subsubsection{Slow-wave bursting model}
We will not enter into details with regards to the simulation of slow-wave bursters, but note that Eq.~\eqref{model_eqs} can be used to produce them if a two-dimensional self-oscillating slow subsystem is used, that is $z\in\mathbb{R}^m,m=2$. Equations \ref{eq_line} and \ref{eq_arc} have to be substituted with appropriate parametrizations of the closed paths shown in Fig.~\ref{fig_focus_bifdiagr_sw}.

\subsection{Summary for the codim-3 deg TB, focus case}

\begin{figure}[t]
\center
\includegraphics[width=0.45\textwidth]{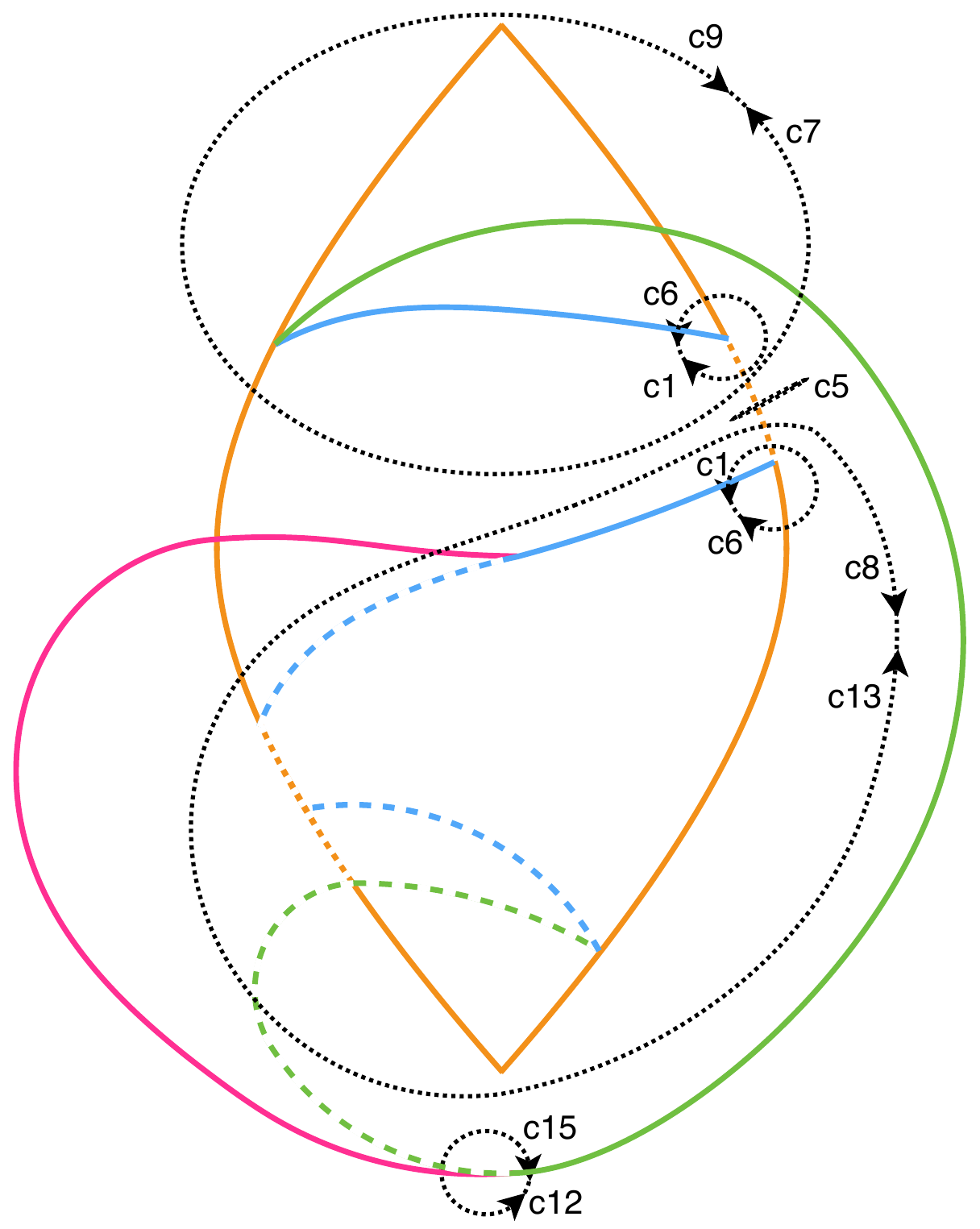}
\includegraphics[width=0.45\linewidth]{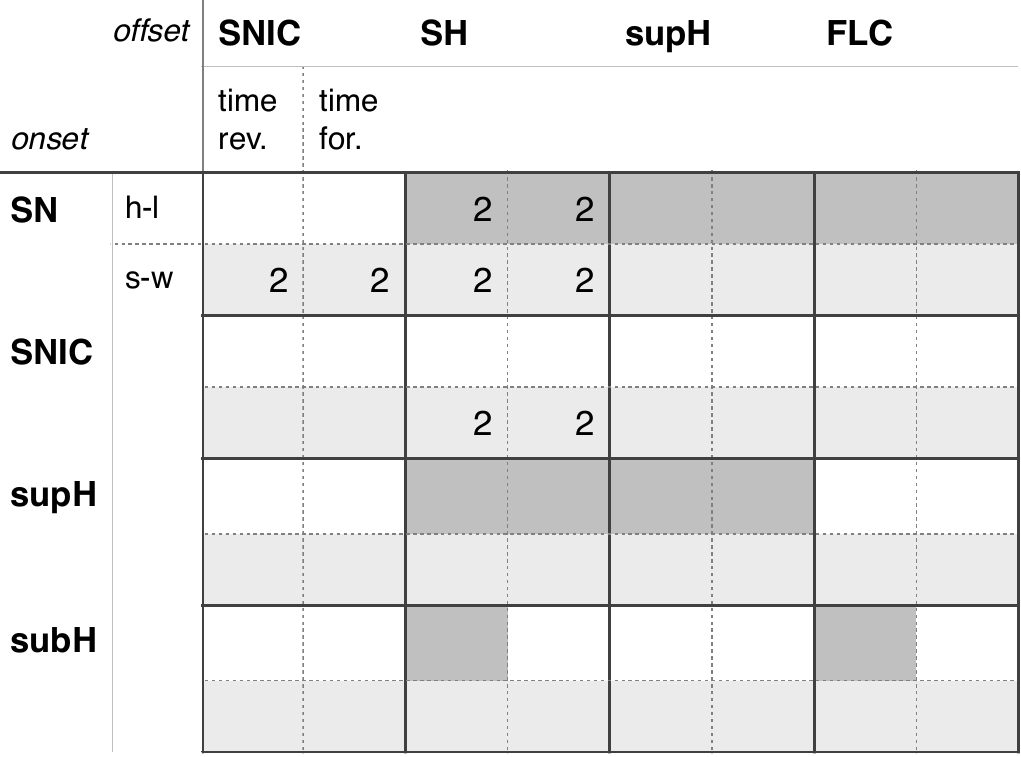}
\caption{\textbf{Focus slow-wave bursters.} In the top panel we show examples of closed paths for the slow-wave bursting classes (dotted black curves). Paths for classes for which an hysteresis-loop counterpart exists are not shown, the hysteresis-loop path can, in fact, be used also for slow-wave bursting.  The table summarizes the results for the deg. TB singularity, focus case. For each class we distinguish (sub-columns) the cases in which the equations of the unfolding are used in time forward or time reversal behavior. In dark grey we mark the existence of hysteresis-loop (h-l) bursting paths for a given class and in light grey that of slow-wave (s-w) bursting paths. The number 2 inside a cell indicates that for that class two realizations in different regions of the unfolding are possible. Overall the highest amount of classes is obtained in the time reversal case: all the sixteen slow-wave are in there and seven hysteresis-loop ones (one of them, c2, with two realizations).}
\label{fig_focus_bifdiagr_sw} 
\end{figure}

Overall, in the unfolding of the codim-3 Takens-Bogdanov bifurcation of focus type, time-reversal, we found seven classes of hysteresis-loop bursters (one of them with two different realizations) and all the sixteen classes of slow-wave bursters. These results are summarized in the table in Fig.~\ref{fig_focus_bifdiagr_sw}. In the table, the results for a similar analysis conducted in the time-forward case are also reported.

\subsection{Classes stability far from the codim-3 singularity}

\begin{figure}
\center
\includegraphics[width=\textwidth]{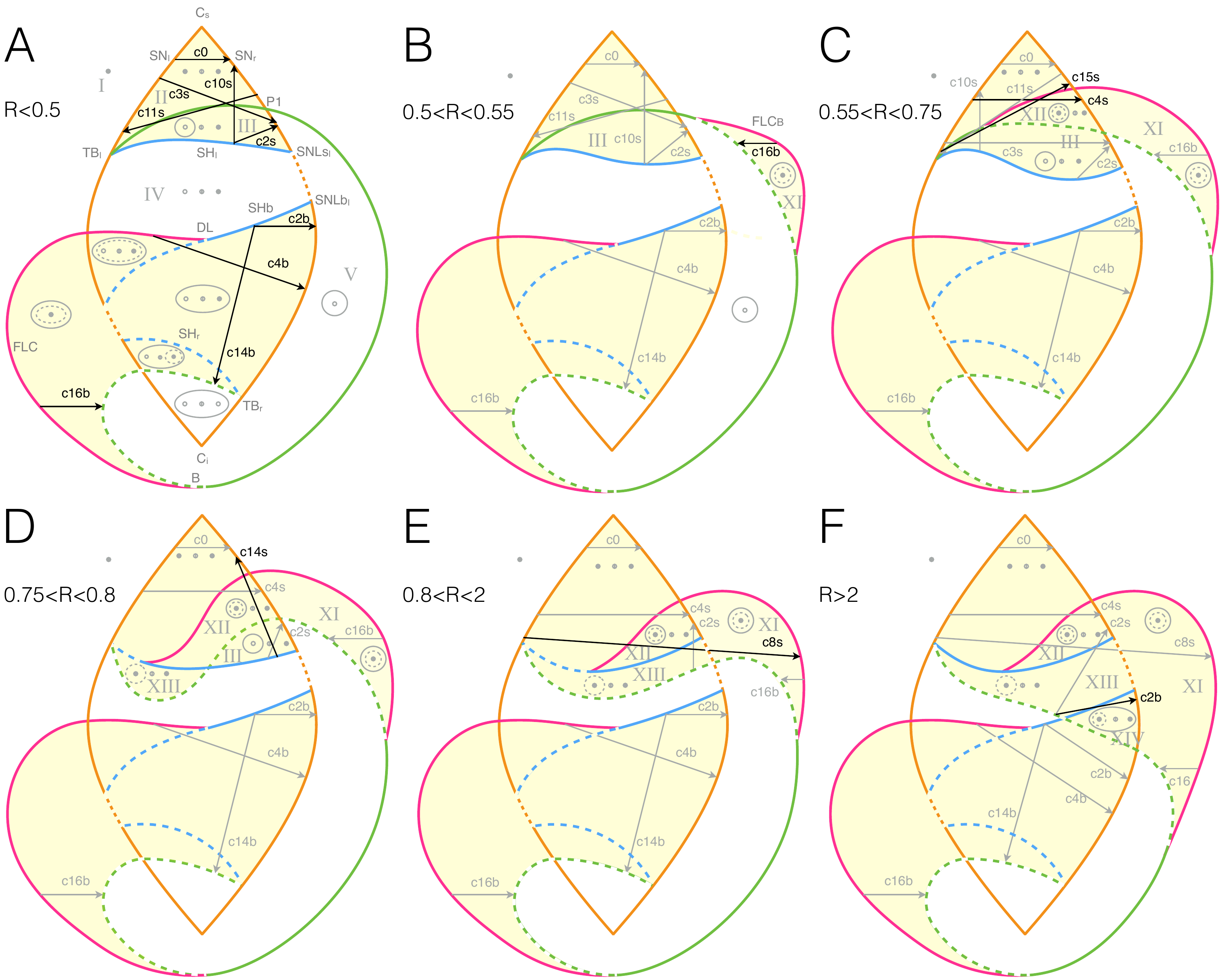}\vspace{.5cm} 
\includegraphics[width=0.7\textwidth]{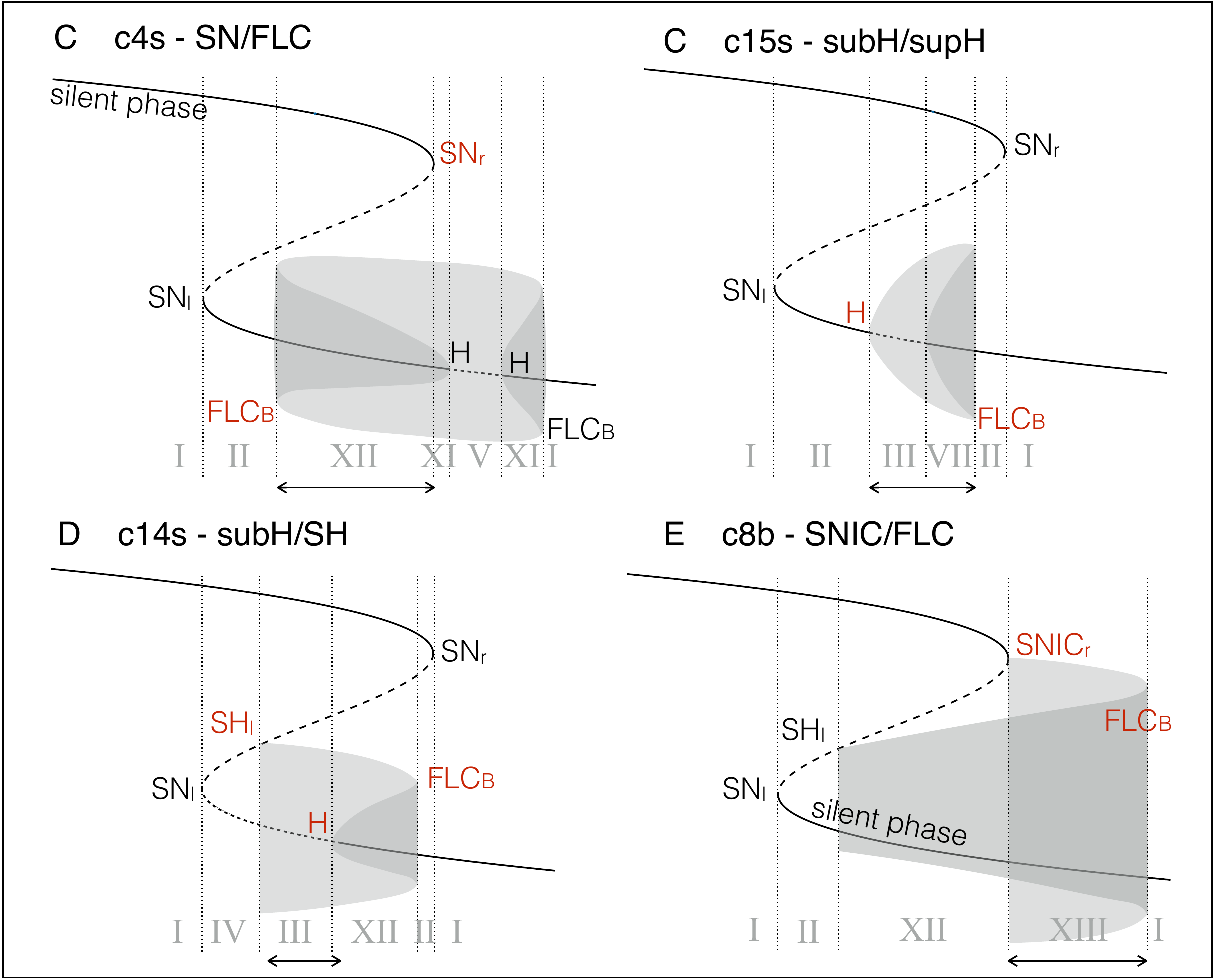}
\caption{\textbf{Effect of increasing the radius of the sphere.} Bifurcation diagrams on spheres of different radius are topologically equivalent if $R<0.5$ (A). For bigger values of the radius, different topologies of bifurcation curves appear as shown in B-E. For each of these stages we verified the existence of bursting paths. Paths for classes already existing for smaller values of R are shown as grey arrows, paths for new classes as black arrows. In the bottom panel we show bifurcation diagrams for the new classes found when increasing R.}
\label{fig_forthparam}
\end{figure}

\textbf{Unfolding far from the codim-3 singularity.} We have discussed in Section \ref{subsection_focus} the topology of the bifurcation diagram produced by the intersection of a sphere centered in the codim-3 singularity and the unfolding of this singularity. Bifurcation diagrams obtained for different values of the radius of the sphere are topologically equivalent provided the radius is small enough \cite{dumortier1991bifurcations}. We investigated numerically the impact of the radius and found that the topology stays the same for $R<0.5$. For increasing values of $R$ one can observe a sequence of topologies as represented in Fig.~\ref{fig_forthparam} (see also \cite{krauskopfcodimension}). In this figure we labeled different stages of topological equivalence with letters from A to F. The upper region of the bifurcation diagram is the most affected by changes in R. At first (B) a new curve of fold limit cycle bifurcation, $FLC_{B}$, separates from the $H$ curve in the part of the unfolding where only a fixed point exists. This curve later (C) crosses $SN_l$ and enters the region where three fixed points coexists. The next topological changes are due to the behavior of the $H$ bifurcation curve in the upper part of the diagram. This curve gradually passes below $SH_l$ (D-E) until it intersects $SHb$ (F). We evaluated the value of the radius for each stage with a precision of $0.05$. Note that all the results in this paper are obtained for $R=0.4$.\\

\textbf{Classes far from the codim-3 singularity.} Do classes, which were found for small values of the radius, survive far from the codim-3 singularity? We examined each of the bifurcation diagrams in Fig.~\ref{fig_forthparam} looking for paths for bursting activity. We observed that some of the classes persist through all the values of $R$ analyzed: they are all the classes in LCb (c2b, c16b, c4b, c14b), plus c0 and c2s in LCs. The other classes found for a small radius (in A) disappear after C. On the other hand, there are some classes that arise for bigger values of the radius, they are classes c4s, c8b, c14s and c15s. While two of them, c4 and c14, already appeared with a different realization (i.e. c4b and c14b) for $R<0.5$, the other two, c8b and c14s, were not found before. Bifurcation diagrams for the new bursting classes are shown in the lower panel of Fig.~\ref{fig_forthparam}. It is worth noticing that c8b, the only class with SNIC bifurcation, has an anomalous bifurcation diagram as compared to all the other classes investigated: in this case the lower branch of fixed points, and not the upper branch, plays the role of silent state.

\subsection{Deg. TB singularity - Elliptic, saddle and cusp cases}
\label{subsection_othercases}

\begin{figure}[t]
\center
\includegraphics[width=0.9\textwidth]{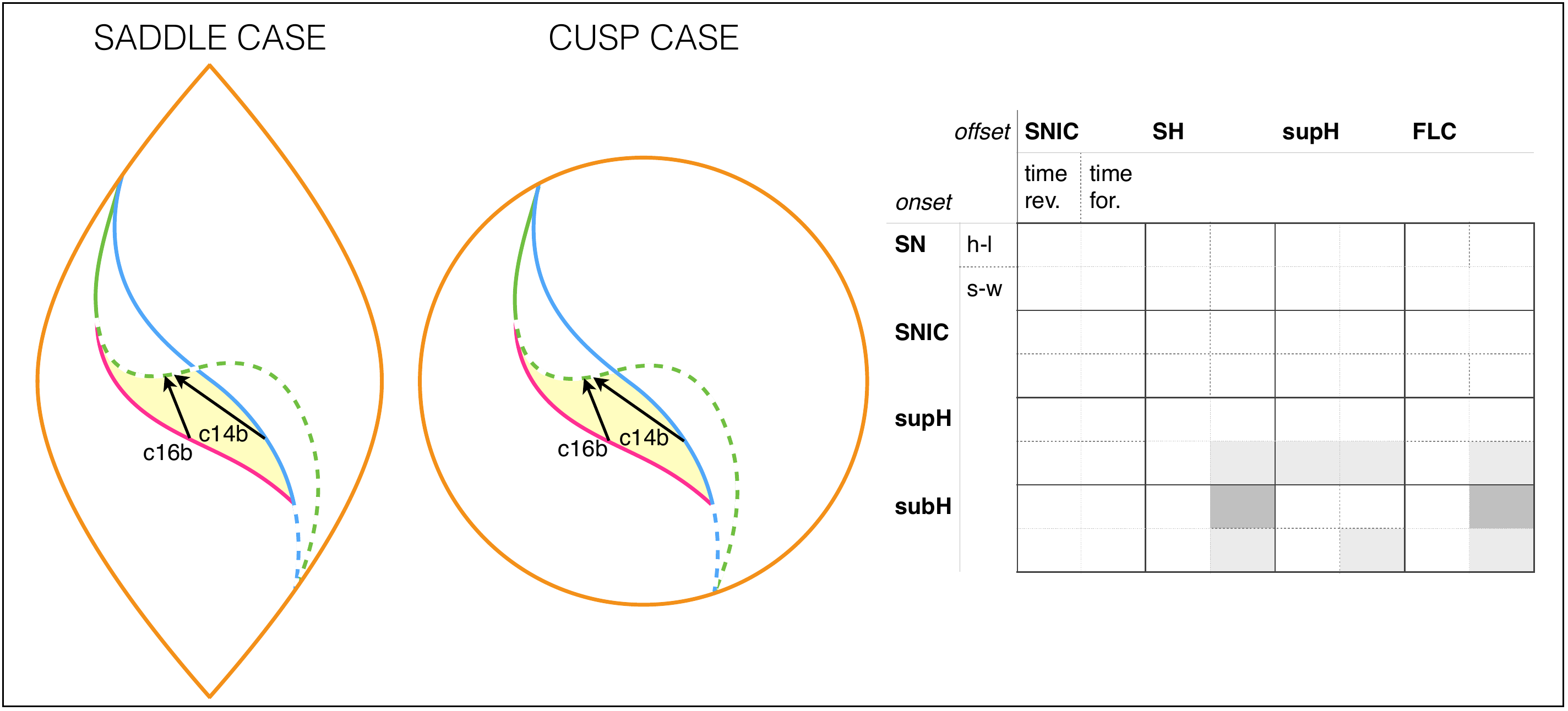}
\caption{\textbf{Deg. TB singularity: saddle and cusp cases}. We show the paths in the unfolding of the saddle case (time-forward) and the overall results in the table.  Paths in the unfolding of the cusp case (time-reversal) are also shown in the bottom panel, the overall results are identical to those of the saddle case, where time-forward and time-reversal are inverted, we thus did not report them in a extra table. The legend is as in in Fig.~\ref{fig_focus_bifdiagr}.}
\label{fig_othercases}
\end{figure}

The results described up to now pertain to the deg. TB focus case. We similarly investigated the unfoldings of the other three cases, the elliptic, saddle and cusp deg. TB singularities \cite{dumortier1987generic,dumortier1991bifurcations,baer2006multiparametric}. This did not add any new class with regards to those located in the focus case.\\

\textbf{Elliptic.} The elliptic case is described by Eq.~\eqref{eq_focus} as well, but this time $b>2\sqrt{2}$. Baer and colleagues \cite{baer2006multiparametric} showed that the description of this unfolding is topologically equivalent to that of the focus case. We thus have the same bursting classes than in the latter. The authors pointed out that in the elliptic case the small limit cycle displays a `canard-like' behavior close to the saddle-node curves, with rapid changes in the amplitude. This renders the numerical continuation of the limit cycle harder than in the focus case. Another difference underlined by Baer and colleagues is that in the elliptic case, the orbit at $SHb$ tends to the boundary of the elliptic sector rather than to the origin when approaching the codim-3 singularity.\\

\textbf{Saddle.} The saddle case is obtained from Eq.~\eqref{eq_focus} when the term $x^3$ has negative sign and for every $b$. In this case two of the three fixed points are saddles, which reduces the possibility of having bistability. We find, in fact, only two hysteresis-loop bursters and five slow-wave ones. Results are summarized in the table in the bottom panel of Fig.~\ref{fig_othercases} and examples of hysteresis-loop paths for the time forward behavior are shown in the left bifurcation diagram.\\

\textbf{Cusp.} The equations for the cusp case are different than for the other cases \cite{dumortier1987generic} (see Supplementary Materials \ref{SM_cusp}). Unlike the other cases the cusp case only allows for two fixed points: one saddle and one focus. Bursting classesare the same as for the saddle case, but the time forward and time reversal behaviors are inverted. Examples of hysteresis-loop paths for the time reversal behavior are shown in the second bifurcation diagram in the bottom panel of Fig.~\ref{fig_othercases}.
\begin{wrapfigure}{t}{0.37\textwidth}
\center
\includegraphics[width=0.32\textwidth]{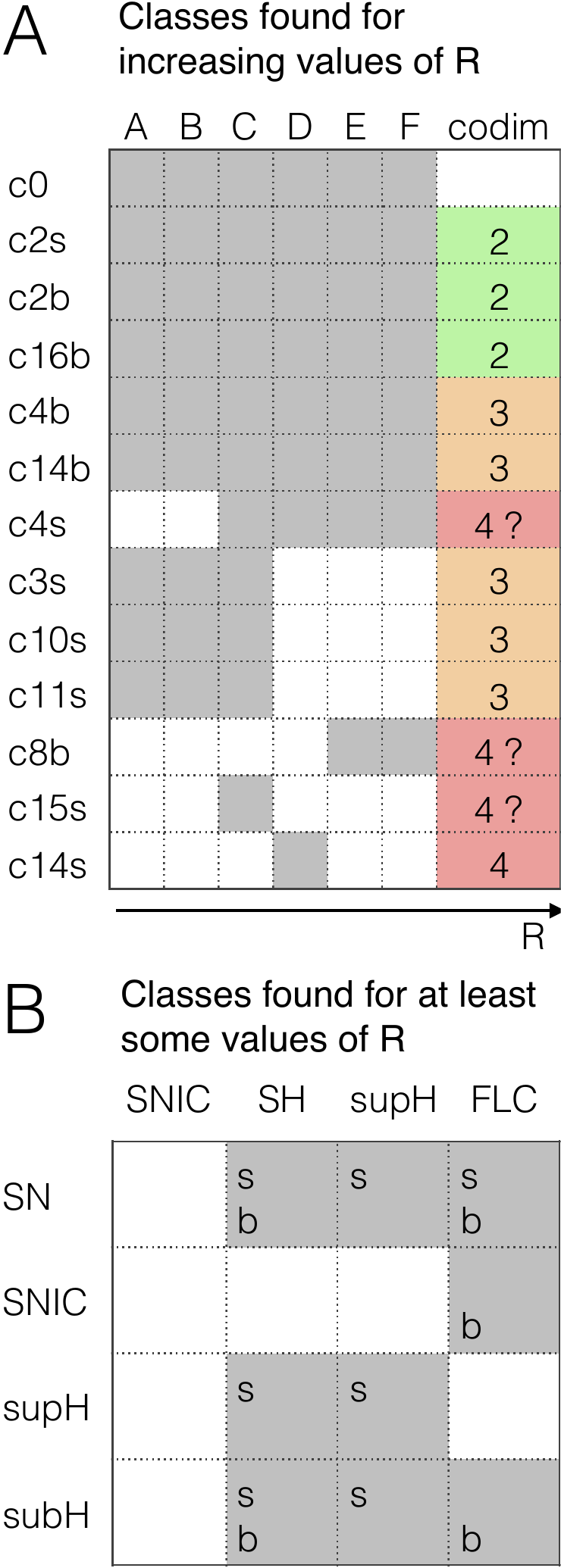}
\caption{\textbf{Existence and complexity of classes.} (A) The letters from A to F refer to the stages of topological equivalence identified in Fig.~\ref{fig_forthparam} for increasing $R$. Some of the classes that were located for small $R$ disappeared far from the codim-3 singularity, while new classes appeared. The figure shows the lifespan of each class as a function of $R$ (grey indicates the existence of a class). For each class we report the smallest codimension of the singularity in which unfolding the class first appears. This can be used as a measure of complexity. We identified hysteresis-loop bursters of complexity 2,3 and 4 (green, orange and red). For the classes with a question mark, further investigation is needed to establish their complexity. (B) The table summarizes the hysteresis-loop bursting classes present in our model, considering also changes in $R$. The letters `s' and `b' indicate whether the classes found belong to the LCs or LCb regions.}
\label{fig_complexity}
\vspace{-1.5cm}
\end{wrapfigure}

\section{Discussion}

\subsection{A unifying framework for fast-slow bursters}

Here we have shown how to build fast-slow planar bursters by appplying the unfolding theory approach to the unfolding of the codim-3 deg. TB singularity. We systematically explored the four possible unfoldings of this planar singularity: namely the focus, elliptic, saddle and cusp case. For each of them we checked the time-forward and time-reversal behavior and located paths for slow-wave and hysteresis-loop bursters. We discovered that the focus case with time-reversal behavior covers the largest number of bursting classes. The other three cases did not bear new classes. In the unfolding of the focus case we found all the sixteen classes for slow-wave bursting and seven of the hysteresis-loop classes predicted by Izhikevich \cite{izhikevich2000neural}. We could identify two additional hysteresis-loop classes when exploring the unfolding far from the singularity.

When increasing the radius of the system, while some additional classes can appear, others are destroyed. The lifespan of each class, as a function of $R$ is reported in Fig.~\ref{fig_forthparam}A. Fig.~\ref{fig_forthparam}B summarizes the hysteresis-loop classes found in the unfolding, considering those located far from the codim-3 singularity. Overall, nine classes out of sixteen are present, some with a realization both in the LCs and in the LCb regions. We could not find c12 and all the classes with SNIC onset or offset, except one.  

We built a model that can produce bursting activity for all the hysteresis-loop classes found. The equations of the fast subsystem of the model are those of the unfolding of the focus case of the deg. TB singularity. The unfolding parameters are parametrized in terms of one slow variable to follow the required path. The slow variable shows oscillations thanks to a linear feedback from the fast subsystem. The system is globally stable, so every initial condition will lead to the same dynamics if the system is in the bursting region. The only parameters that affect whether bursting is present or not, and determine the bursting class, are the starting and ending points of the path. All the other parameters of the model can affect the shape of the orbit followed by the system without changing the class. For example, we showed how they can increase/decrease the number of oscillations in the active phase, alter their amplitude and frequency (within the constraints imposed by the bifurcations involved), or modify the silent/active phases lengths ratio. 

This provides a unifying framework to investigate the underlying mechanisms of systems able to display different bursting behaviors. Furthermore, transitions between classes can be easily implemented through an ultra-slow modulation of initial and final points of the path.

We could make predictions about which transitions are easier to obtain and which instead require to cross regions of the unfolding where the system is not bursting. We also clarified why transitions are easier to obtain for hysteresis-loop bursters than for slow-wave ones.

\subsection{Finding new paths for bursting}

Bertram et al. \cite{bertram1995topological} searched for bursting paths in the two parameters bifurcation diagram of the Chay-Cook model. This bifurcation diagram, as pointed out by the authors, is a layer of the unfolding of the deg. TB singularity, focus case. It can be obtained by keeping $\mu_2$ fixed at a negative value. Such a layer describes a bifurcation diagram that excludes some of the points that we find on the sphere: the two cusp points $C_s,C_i$ (they require $\mu_2=0$), and the Bautin point (requires a positive $\mu2$).  In this layer, Bertram and colleagues identified c2s, c2b, c4b and c16b with hysteresis, and c5 without hysteresis. De Vries \cite{de1998multiple} added the path for c3s with hysteresis. The complete unfolding on the sphere has been investigated by Osinga et al. \cite{osinga2012cross}. The authors located paths for the bursters known to Bertram et al. and de Vries and they added the path for c10s and. By exploring the unfolding far from the cosim-3 singularity they also added a new class. This extra class is the point-point SN/subH burster. Izhikevich \cite{izhikevich2000neural} and Goulubitsky et al. \cite{golubitsky2001unfolding} identified additional slow-wave classes close to two codim-2 bifurcations that exists also in the unfolding of the deg. TB singularity: the SNL point (c1,c6) and the B point (c11,c12,c15). In this article we have provided a systematic framework of bursting activity, located paths for bursting in parameter space and identified novel paths for c4b, c8, c11, c12, c14 and c15 for hysteresis-loop and all the missing classes for slow-wave.

\subsection{Complexity of bursting classes}

Golubitsky et al. \cite{golubitsky2001unfolding} introduced a notion of \textit{complexity} in the characterization of bursters. They define the complexity of a bursting class as the codimension of the lowest codimension singularity in which unfolding that class can be found. Onset and offset bifurcations, in facts, are not enough to describe the sequence of bifurcations required by a bursting class. Other bifurcations may be required to obtain a class and even a greater number of them may be needed for the hysteresis-loop types. If more bifurcations are required for a class, then more parameters must be tuned to obtain a given sequence. This increases the complexity of the class. The number of parameters to be tuned is reflected in the codimension of the singularity in which unfolding the class first appears. Thus, Golubitsky and colleagues argued that the more complex a class is, the less likely it is to encounter that class, in empirical data or in models. They propose to complement Izhikevich's classification by providing information on this measure of complexity.

With regards to hysteresis-loop bursters, the classes of smallest complexity are c2 and c16, which exist close to codim-2 bifurcations (Saddle-Node-Loop and Bautin bifurcations respectively) as outlined by Golubitsky et al. Fig.~\ref{fig_forthparam} reports the complexity of the classes identified in the present work. We classified the classes appearing far from the codim-3 singularity as codim-4 classes. With regards to these classes, it is an open question whether they could be located in the proximity of a codim-3 singularity other than the deg. TB. As pointed out by Krauskopf et al. \cite{osinga2012cross}, not all the unfoldings of codim-3 singularities are available. Nonetheless, the authors noted that the unfolding of the deg. TB is the only one to present both a codim-3 cusp point and a codim-2 TB bifurcation (at which Hopf and saddle-homoclinic bifurcations coincide). From this we can conclude that classes that require all these conditions, such as c14s, cannot be found in the unfolding of other codim-3 singularities. Thus, their complexity is four. Further work is required to determine the codimension of classes c8b, c15s, c4s.

In the present work, we focused on point-cycle bursters, that is bursters with a fixed point like silent phase and with a limit cycle for the active phase. We also provide an example of a point-point burster, in which both phases are given by fixed points. Other possibilities exist that we did not address, as for example cycle-cycle bursters, in which the silent phase is characterized by small amplitude oscillations, requiring the coexistence of two stable-limit cycles. 

\subsection{Alternative modeling approaches}

Franci et al. \cite{franci2014modeling} applied the unfolding theory approach to a codim-3 singularity to build a three variables model for bursting activity. 
They used the unfolding of the codim-3 winged cusp singularity, which is described by a single variable $x$. The unfolding presents regions with only  one fixed point and regions with coexistence of three fixed points. No limit cycle can live in one dimension. To generate oscillations Franci and coworkers expressed the unfolding parameters in terms of a second variable $y$ acting on a slower time-scale. The second variable receives feedback from the fast one and exploits the hysteresis present in the fast variable to create a limit cycle in the plane $( x,y )$. This second variables plays the same role as our $z$ in class c0 in Fig.~\ref{fig_focus_bifdiagr}: no limit cycle is present in the fast subsystem but, thanks to hysteresis, one can be created in the $( x,y )$ plane (note that the spiraling towards the fixed point in c0 is due to the presence of the second fast variable and is thus absent in the limit cycle generated by Franci et al. in the $( x,y )$ plane). The authors introduced a third variable, $z$, acting on a slower time-scale than $y$, that allows the alternation between active and silent phases with a similar mechanism of that used here. By changing the parameters they provided a model for c2, c3, c4 and c16 hysteresis-loop bursters. They also showed an example of how an ultra-slow modulation can lead to transitions between classes.

In contrast, our approach is based upon the planar unfolding of the codim-3 deg. TB singularity. With the same number of variables we identified a richer repertoire of bursting activity. This is due to the fact that the planar unfolding of the codim-3 deg. TB is richer, in terms of limit cycles and bifurcations, than the one-dimensional unfolding of the codim-3 winged cusp coupled with a slow variable.

The unfolding theory approach proves to be a valuable tool to build bursters when timescale separation holds. When this does not happen, then different phenomena can occur, including chaos \cite{izhikevich2000neural}.

In our study bursting behavior was engineered by describing slow changing parameters. However, this is not the only way of eliciting bursts. Both slow and fast interventions can provoke bursts. For parameter changes the changes are usually slow (as discussed throughout the paper). However, the state variables are sensitive to perturbations, for example, because of multi-stability. In this study, the limit cycle that encircles all fixed points is less sensitive to perturbations than the limit cycle that does not encircle all fixed points (see Fig.~\ref{fig_flows}). The effect of perturbations is mostly reversible in the latter case, where in the former case the fast event needs to be coordinated (e.g. temporally) to reverse an effect of perturbations (see Spiegler et al., \cite{spiegler2010bifurcation}, for example Figures 10-12). The existence of a ‘small’ LC next to others or fixed points can be directly used to describe bursting behavior by fast events, for instance, in repetitive spiking sequences. Other dynamics associated with bursting behavior are quasi-periodicity, deterministic chaos and intermittency \cite{gu2014difference}.

\section{Supplementary Materials}

\subsection{Cusp case - equations.} 
\label{SM_cusp}
Equations for the the unfolding of deg. TB singularity, cusp case are given by \cite{dumortier1987generic}:

\begin{equation}
\begin{cases}
\dot{x}={y}\\
\dot{y}=x^2+\mu+y(\nu_0+\nu_1x\pm x^3)\\
\end{cases}
\end{equation}

\printbibliography

\end{document}